\documentclass[]{pasj01}  
\begin{document}   
\Received{} \Accepted{}   
\title{Rotation and Mass in the Milky Way and Spiral Galaxies}
\author{Yoshiaki \textsc{Sofue}\altaffilmark{1}
}
\altaffiltext{1}{Institute of Astronomy, The University of Tokyo, Mitaka, 181-0015 Tokyo } 
\email{sofue@ioa.s.u-tokyo.ac.jp}
\maketitle      

\def\r{\bibitem[]{}}  
\def\be{\begin{equation}} \def\ee{\end{equation}} \def\bc{\begin{center}} 
\def\ec{\end{center}} \def\bfig{\begin{figure}}   \def\ef{\end{figure}}  
\def\bt{\begin{table}}   \def\et{\end{table}}  \def\cen{\centerline}  
 \def\Msun{ M_{\odot \hskip-5.2pt \bullet} } 
\def\msun{ $M_{\odot \hskip-5.2pt \bullet}$ } 
\def\msqpc{\Msun {\rm pc}^{-2}} \def\mcupc{\Msun {\rm pc}^{-3}}  
\def\Vsun{ V_0 } \def\Rsun{ R_0 } \def\rzero{ $R_0$ } \def\vzero{ $V_0$ }  
\def\w{\omega} 	\def\wsun{\omega_0} \def\dv{ de Vaucouleurs } 
\def\gc{ Galactic Center } \def\kms{km s$^{-1}$ }  
\def\kmsmpc{ km s$^{-1}$ Mpc$^{-1}$ } 
\def\kmskpc{$ {\rm km~s^{-1}~kpc^{-1}}$ } 
\def\vr{v_{\rm r}} \def\vp{v_{\rm p}} \def\vrot{ $V_{\rm rot}$ } 
\def\Vrot{ V_{\rm rot} }  \def\Vrotr{V_{\rm rot}^{v_{\rm r}}} 
\def\Vrotp{V_{\rm rot}^\mu} \def\Vrotv{V_{\rm rot}^{\rm vec}}
\def\ha{H$\alpha$\ } 	\def\Ha{ H$\alpha$\ } \def\halpha{ H$\alpha$\ } 
\def\Halpha{ H$\alpha$ }  \def\Deg{^\circ} \def\deg{^\circ} 
 \def\lv{(l, v)}   \def\cos{~{\rm cos}~} \def\sin{~{\rm sin}~} 
\def\sinl{~{\rm sin~}l~} \def\cosl{~{\rm cos~}l~}  \def\masy{ mas y$^{-1}$ }
\def\vp{v_{\rm p}} \def\vr{v_{\rm r}} \def\Up{U_{\rm p}} 
\def\Ur{U_{\rm r}} \def\V0{ V_0 }
\def\Mb{M_{\rm b}} \def\Md{M_{\rm d}} \def\Mbd{M_{\rm b+d}}
\def\Mbdh{M_{\rm b+d+h}} \def\Mh{M_{\rm h}} 
\def\ab{a_{\rm b}} \def\ad{a_{\rm d}} \def\ah{a_{\rm h}} 
\def\bc{\begin{center}} \def\ec{\end{center}} 
\def\kms{ km s$^{-1}$ } \def\Msun{M_\odot}
\def\be{ \begin{equation}} \def\ee{\end{equation}  }
\def\ab{a_{\rm b}} \def\ad{a_{\rm d}} \def\Mb{M_{\rm b}} 
\def\Md{M_{\rm d}} 
\def\mb{M_{\rm b}} \def\md{M_{\rm d}} \def\Mh{M_{\rm h}} 
\def\Mbd{M_{\rm b+d}} \def\mbd{M_{\rm b+d}} 
\def\mh{M_{\rm h}} \def\mhalo{M_{200}} \def\rhalo{R_{200}}  
\def\mbdh{M_{\rm 200+b+d}} \def\sb{\rm SMD_b} \def\sd{\rm SMD_d}  
\def\dv{ de Vaucouleurs } \def\ha{H$\alpha$}
\def\mtwelve{10^{12}\Msun} \def\meleven{10^{11}\Msun} 
\def\mten{10^{10}\Msun} \def\log{{\rm log}_{10}} \def\amh{M_{\rm h}}
\def\changed{ }
     
\begin{abstract} 
Rotation curves are the basic tool for deriving the distribution of mass in spiral galaxies. In this review, we describe various methods to measure rotation curves in the Milky Way and spiral galaxies. We then describe two major methods to calculate the mass distribution using the rotation curve. By the direct method, the mass is calculated from rotation velocities without employing mass models. By the decomposition method, the rotation curve is deconvolved into multiple mass components by model fitting assuming a black hole, bulge, exponential disk and dark halo. The decomposition is useful for statistical correlation analyses among the dynamical parameters of the mass components. We also review recent observations and derived results.\\
\\
Full resolution copy is available at URL:\\
http://www.ioa.s.u-tokyo.ac.jp/$\sim$sofue/htdocs/PASJreview2016/
\\
\end{abstract}    

		\chapter{INTRODUCTION}
  
Rotation of spiral galaxies is measured by spectroscopic observations of emission lines such as H$\alpha$, HI and CO lines from disk objects, namely population I objects and interstellar gases. In these lines the velocity dispersion is negligibly small compared to rotational velocity, which implies that the pressure term is negligible in the Virial theorem, so that the dynamical balance between the gravitational and centrifugal forces may be used to calculate the mass in sufficient accuracy.  Absorption lines are also used for bulge's kinematics using velocity dispersion and rotation. The dynamical approach is essential particularly for measurement of the mass of the dark matter and black holes, which are not measurable in surface photometry assuming a mass-to-luminosity ratio.  

In sections 2 and 3 we review the various methods to derive rotation curves of the Milky Way and spiral galaxies, respectively, and describe the general characteristics of observed rotation curves. The progress in the rotation curve studies will be also reviewed briefly.  
In section 4 we review the methods to determine the mass distributions in disk galaxies using the rotation curves, and describe their dynamical mass structures. 

By definition, a rotation curve is the mean circular velocity around the nucleus as a single function of radius. Non-circular streaming motions such as due to spiral arms, bars, and/or expansion/contraction motions are not considered here. Limitation of the current rotation curve analyses is discussed in section 2.
 Elliptical galaxies, for which rotation curve analysis is not applicable, are beyond the scope of this review. Considerations that employ unconventional physical laws such as MOND (modified Newtonian dynamics) are also out of the scope of this review. 

There are a number of articles and reviews on rotation curves and mass determination of galaxies, that include Sofue and Rubin (2001) and Sofue (2013a), and the literature therein. Individual references will be given in the related sections in this review.

  		\chapter{ROTATION OF THE MILKY WAY} 

\section{ Dynamical Parameters Representing the Galaxy} 

In this review, we use either the traditional galactic constants of  $(R_0, V_0)$=(8.0 kpc, 200 \kms), or the recently determined values, $(R_0, V_0)$=(8.0 kpc, 238 \kms) from observations using VERA (VLBI Experiment for Radio Astrometry) (Honma et al. 2012; 2015). Here, \rzero is the distance of the Sun from the Galactic Center and \vzero is the circular velocity of the Local Standard of Rest (LSR) at the Sun (see Fich and Tremaine 1991 for review).

An approximate estimation of the mass inside the solar circle can be obtained for a set of parameters of \rzero = 8 kpc and \vzero = 200 to 238 \kms, assuming spherical distribution of mass, as
\be
M_0={R_0 V_0^2 \over G} = (7.44 {\rm ~to~} 1.05)\times 10^{10} \Msun \sim 10^{11}\Msun,
\label{eq-masssolcirc}
\ee 
with $G$ being the gravitational constant, and the Solar rotation velocity \vzero being  related to \rzero as
$V_0=(A-B)R_0$,  $A$ and $B$ are the Oort's constants, which are determined by measuring radial velocity and proper motion of a nearby star. 
See Kerr and Lynden-Bell (1986) for a review about the Oort constants, and tables \ref{tabAB} and \ref{tabR0V0} for recent values.

\begin{table*} 
\caption{Oort's constant $A$ and $B$. } 
\begin{center}
\begin{tabular}{lll}
\hline
Authors (year) & $A$ \kmskpc & $B$ \kmskpc \\ 
\hline 
IAU recommended values (1982)&$ 14.4\pm 1.2$ &$ -12.0 \pm 2.8 $\\
HIPPARCOS (Feast and Whitelock 1997)&$ 14.8\pm 0.8$ &$ -12.0\pm 0.6 $\\
Cepheids (Dehnen and Binney 1997)&  $ 14.5 \pm 1.5$ & $ -12.5 \pm 2$ \\ 
Dwarfs, infrared photometry (Mignard 2000)& $11.0\pm 1.0$ & $-13.2\pm 0.5$\\
---- Distant giants& $14.5\pm 1.0$ & $-11.5\pm 0.5$\\
Stellar parallax (Olling and Dehnen 2003) & $\simeq 16$ & $\simeq -17$ \\
\hline
\end{tabular}   
\end{center}
\label{tabAB}
\end{table*}

\begin{table*} 
\caption{Galactic constants ($R_0,V_0$). } 
\begin{center}
\begin{tabular}{lll}
\hline
Authors (year) & $R_0$ kpc & $V_0$ \kms \\ 
\hline 
IAU recommended (1982)& 8.2  & 220  \\
Review before 1993 (Reid 1993) &$8.0 \pm 0.5$   &  \\
Olling and Dehnen 1999 & $7.1\pm 0.4$ & $184\pm 8$ \\
VLBI Sgr A$^*$ (Ghez et al. 2008)&$8.4 \pm 0.4$  &  \\
ibid (Gillessen et al. 2009)& $8.33 \pm 0.35$  &  \\
Maser astrometry (Reid et al. 2009) & $8.4\pm 0.6$  & $254\pm 16$  \\
Cepheids (Matsunaga et al. 2009) & $8.24 \pm 0.42$ \\
VERA (Honma et al. 2012, 2015) & $8.05\pm 0.45$ & $238\pm 14$ \\
\hline
\end{tabular}   
\end{center}
\label{tabR0V0}
\end{table*}

By the definition of rotation curve, we assume that the motion of gas and stars in the Galaxy is circular. This assumption put significant limitation on the obtained results. In fact, the galactic motion is superposed by non-circular streams such as a flow due to a bar, spiral arms, and expanding rings. The dynamical parameters of the Galaxy to be determined from observations, therefore, include those caused both by axisymmetric and non-axisymmetric structures. In table \ref{parameters} we list the representative parameters and analysis methods (Sofue 2013b). 

In the present paper, we review the methods to obtain parameters (1) to (10) in the table, which define the axisymmetric structure of the Galaxy as the first approximation to the fundamental galactic structure.
Non-circular motions caused by a bar and spiral arms are beyond the scope of this review, which are reviewed in the literature (e.g., Binney et al. 1991; Jenkins and Binney 1994; Athanassoula 1992; Burton and Liszt 1993). 

\begin{table*}[htbp]
\caption{Dynamical parameters for the Galactic mass determination$^\dagger$ (Sofue 2013b).}
\begin{center} 
\begin{tabular}{llll}
\hline \hline
Subject &  Component & No. of Parameters \\ 
\hline
I. Axisymm. &  Black hole  &(1) Mass \\
structure &  Bulge(s)$^\ddagger$ &  (2) Mass \\
(RC) && (3) Radius\\
&&(4) Profile (function)\\
&Disk& (5) Mass\\
&&(6) Radius \\
&&(7) Profile (function)\\
&Dark halo& (8) Mass\\
&&(9) Scale radius\\
&&(10) Profile (function)\\
\hline 
II. Non-axisymm. &Bar(s)& (11) Mass \\
structure&& (12) Maj. axis length\\
(out of RC)&& (13) Min. axis length\\
&& (14) $z$-axis length\\
&& (15) Maj. axis profile\\
&& (16) Min. axis profile\\
&& (17) $z$-axis profile\\
&& (18) Position angle\\
&&(19) Pattern speed $\Omega_{\rm p}$\\
& Arms& (20) Density amplitude& \\
&&(21) Velocity amplitude\\
&& (22) Pitch angle\\
&& (23) Position angle\\ 
&& (24) Pattern speed $\Omega_{\rm p}$\\
\hline 
III. Radial flow & Rings &(25) Mass\\
(out of RC)&& (26) Velocity\\
&& (27) Radius\\
\hline 
\end{tabular}

$^\dagger$ In the present paper we review on subject I.\\
$^\ddagger$ The bulge and bar may be multiple, increasing the number of parameters.
\end{center}
\label{parameters}
\end{table*}

\section{Progress in the Galactic Rotation Curve}

\begin{figure}
\begin{center} 
\includegraphics[width=6cm]{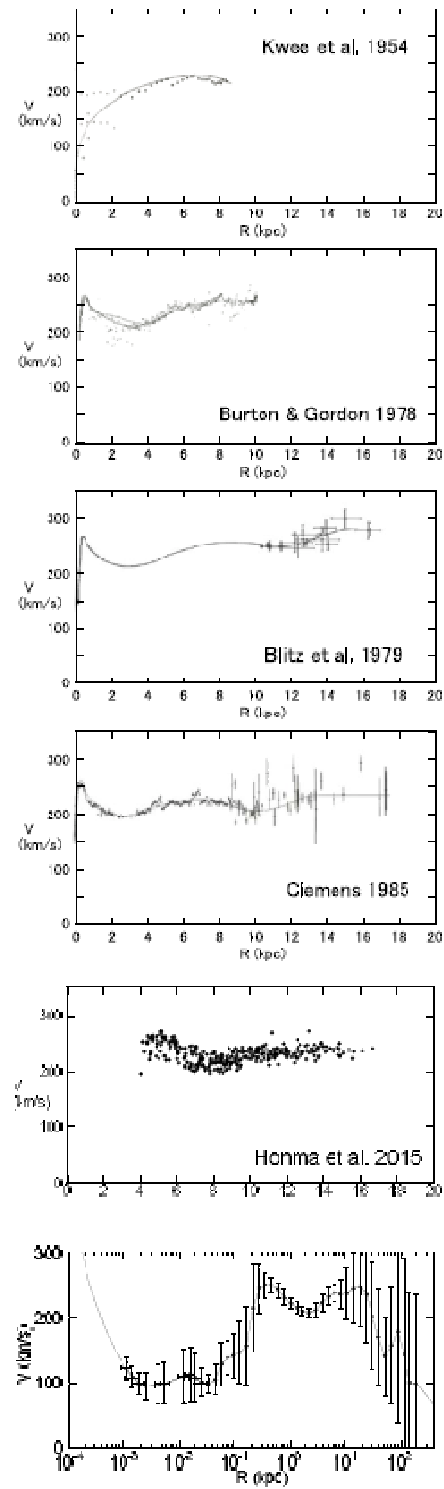} 
\end{center}
\caption{Half a century progress in rotation curve (RC) determination of the Milky Way. 
Top to bottom: 1950's  (Kwee et al. 1954);  
1970's  CO and HI (Burton and Gordon 1978; Blitz et al 1979); 
1980's Composite of CO, HI and optical   (Clemens 1985);
2001's most accurate trigonometric RC by maser sources with VERA (Honma et al. 2015); and a semi-logarithmic grand RC from the Galactic Center, linked to the massive black hole, to half a way to M31 (Sofue 2015a). 
}
\label{progressinmwrc} 
\end{figure}

Figure \ref{progressinmwrc} shows the progress in the determination of rotation curve of the Milky Way Galaxy. Before and till the 1970's the inner rotation curve of the Milky Way has been extensively measured by the terminal-velocity method applied to radio line observations such as the HI line (Burton and Gordon 1978; Clemens 1985; Fich et al. 1989).  

In 1980 to 1990', outer rotation velocities of stellar objects (OB stars) were measured by combining optical distances and CO-line velocities of associating molecular clouds (Blitz et al. 1982; Demers and Battinelli 2007). The HI thickness method was also useful to measure rotation of the entire disk (Merrifield 1992; Honma and Sofue 1997). 

In 1990 to 2000's, VLBI measurements of parallax and proper motions of maser sources and Mira variable stars have provided with advanced rotation velocities with high accuracy (Honma et al. 2007). It was also recent that proper motions of a considerable number of stars were used for rotation curve measurement (Lopez-Corredoira et al. 2014). 

The most powerful tool to date to measure the rotation of the Milky Way up to $R\sim 20$ kpc is the VERA, with which trigonometric determination of both the 3D positions and velocities are done simultaneously for individual maser sources (Honma et al. 2007; 2012; 2015; Sakai et al 2015; Nakanishi et al. 2015).

Since 1990's, with the kinematics of interstellar gas and infrared stars in the vicinity of the nucleus of our Galaxy, Sgr A$^*$, the existence of the massive black hole has been revealed. The black hole mass on the order of $\sim 4\times 10^6 \Msun$ was measured by proper-motion measurements of infrared sources around the nucleus  (Genzel et al. 1994; Ghez et al. 1998; Lindqvist et al. 1992; Gillessen et al. 2009).

For the total mass of the Galaxy including the extended dark halo, analyses of the outermost rotation curve and motions of satellite galaxies orbiting the Galaxy have been obtained in detail. The total mass of the Galaxy including the dark halo up to $\sim 150$ kpc has been estimated to be $\sim 3 \times 10^{11}\Msun$. (Sofue 2015b).

\begin{table*}
\caption{Rotation curves of the Milky Way Galaxy}  
\begin{tabular}{llll}

\hline
Authors (year) & Radii & Method & Remark \\ 
\hline
\\
Burton and Gordon (1978) & 0 - 8 kpc & HI tangent & RC\\
Blitz et al. (1979) & 8 - 18 kpc& OB-CO association & RC \\
Clemens (1985) & 0 -18 kpc & CO/compil. & RC\\
Dehnen  and  Binney (1998) & 8 - 20& compil. + model &RC/Gal. Const.\\
Genzel et al. (1994--), Ghez et al. (1998--), et al.&0--0.1 pc &IR spectroscopy &Orbits, velo. dispersion\\    
Battinelli, et al. (2013) &  9 - 24 kpc& C stars &RC\\
Bhattacharjee et al.(2014) &  0 - 200 kpc& Non-disk objects &RC/model fit \\
Lopez-Corredoira (2014)& 5 - 16 kpc& Red-clump giants $\mu$ & RC\\
Bobylev (2013); --- \& Bajkova (2015)& 5 - 12 kpc & Masers/OB stars & RC/Gal. const. \\
Honma et al. (2012, 2013, 2015)& 3 - 20 kpc& Masers, VLBI & RC/Gal. const.\\
Sofue et al. (2009); Sofue (2013b, 2015a) &0 - 300 kpc & CO/HI/opt/compil. &RC/model fit \\ 
\\ 
\hline
\end{tabular}  
\label{tabrcmw}
\end{table*} 

\section{Methods to Determine the Galactic Rotation Curve}

The radial velocity, $\vr$, and perpendicular velocity to the line of sight, $\vp$, of an object at a distance $r$ orbiting around the Galactic Center are related to the circular orbital rotation velocity $V$  as
\be
\vr= \left({R_0 \over R} V - V_0\right) \sinl,
\label{eq-vr}
\ee
and
\be
\vp=\mu r =-{s \over R} V -V_0 \cosl,
\label{eq-vp}
\ee
where
\be
s=r-R_0 \cosl
\ee
Here $\mu$ is the proper motion, and $R$ is the galacto-centric distance related to $r$ and galactic longitude $l$ by 
\be
R=\sqrt{r^2+R_0^2 - 2 r R_0 \cosl},
\label{Rrl}
\ee 
or
\be
r=R_0 \cosl \pm \sqrt{R^2-R_0^2 {\rm sin}^2 l}
\label{eq-r}.
\ee

\subsection{Terminal-velocity method inside the solar circle}
  
\begin{figure}
\begin{center}  
\includegraphics[width=7cm]{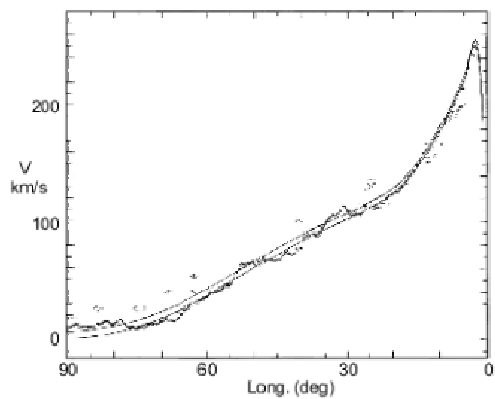}   
\end{center}
\caption{CO and HI line tangent velocities for the inner rotation curve (Burton and Gordon 1978)}
\label{burton1978hitan}
\end{figure}

Inside the solar circle, the galactic disk has tangential points, at which the rotation velocity is parallel to the line of sight. Figure \ref{burton1978hitan} shows the tangent velocities measured for the 1st quadrant of the galactic disk (e.g., Burton and Gordon 1978).  

Maximum radial velocities ${\vr}_{\rm~ max}$ (terminal or the tangent-point velocity) are measured by the edges of spectral profiles in the HI and CO line emissions at $0<l<90\Deg$ and $270<l<360\Deg$. The rotation velocity $V(R)$ is calculated using this velocity as 
\begin{equation}
 V(R) = {\vr}_{\rm ~max} + \Vsun~\sin~ l,
\end{equation}
and the galacto-centric distance is given by 
\be
R=\Rsun \sin~ l.
\ee  
 
\subsection{Radial velocity method} 

The method to calculate the rotation velocity using the radial velocity is applied to various objects, where
\be
 V(R)= {R \over R_0} \left({v_{\rm r} \over \sin l} +V_0 \right). 
\label{vfromvr}
\ee 
In this method, the distance $r$ has to be measured independently of other methods such as trigonometric and/or spectroscopic measurements. Often used objects are OB stars or association. The distances of OB stars are measured from their distance modulus, and the distance is assumed to be the same as that of its associated molecular cloud or HII region whose radial velocity is observed by molecular/recombination line measurements. However, errors in the photometric distances are usually large, resulting in large errors and scatter in the outer rotation curve determinations.

\subsection{Proper motion method}

The rotation velocity can be also measured by using proper motion $\mu$ as
\be
V(R) = - {R \over s} (v_{\rm p} + V_0 \cos l).
\ee  
VLBI techniques have made it possible to employ this method, where the distance $r$, proper motion $\vp$ ($=r\mu$), and radial velocity $\vr$ are observed at the same time.  
Applying this technique to a number of maser sources, the outer rotation curve has been determined with high accuracy (Honma et al. 2012, 2015; Sakai et al. 2012, 2015; Nakanishi et al. 2015) (figure \ref{progressinmwrc}). 

Proper motions of a large number of stars from HYPPARCOS observations combined with the 2MASS photometric data have recently been analyzed for galactic kinematics (Roeser et al 2010; Lopez-Corredoira 2014). Proper motions were obtained from PPMXS catalogue from HYPPARCOS observations, and the distances were determined from K and J band photometry using 2MASS star catalogue correcting for the interstellar extinction. The RCG stars were used for their assumed constant absolute magnitudes.

\subsection{Trigonometric method using velocity vectors with VLBI}

The ultimate method to investigate the Milky Way's rotation, without being bothered by various assumptions such as the circular rotation and/or a priori given solar constants, is the VLBI method, by which three-dimensional (3D) positions and motions of individual maser sources are measured simultaneously.

The VERA observations have most successfully obtained rotation velocities for several hundred galactic maser sources within $\sim 10$ kpc from the Sun. This project has introduced a new era in Galactic astronomy, providing us with the most precise solar constants ever obtained (Honma et al. 2007; 2012; 2015; Sakai et al 2015; Nakanishi et al. 2015). 

When the three independent quantities of $\vr$, $\mu$, and $r$ are known or have been observed simultaneously, the three-dimensional velocity vector can be determined uniquely without making any assumptions such as a circular orbit. The absolute value of the velocity vector is given by
\be
V=\sqrt{\Up^2+ \Ur^2},
\label{eq-Vvpvr}
\ee
where
\be
\Up=\vp+\V0 \cosl
\ee
 and
 \be
 \Ur=\vr + \V0 \sinl.
\ee

\subsection{Ring thickness method}

In the current methods distances of individual objects are measured independently of radial velocities or proper motions. However, distances of diffuse and/or extended interstellar gases are not measurable. 
The HI-disk thickness method has been developed to avoid this inconvenience (Merrifield 1992; Honma and Sofue 1997), where annulus-averaged rotation velocities are determined in the entire disk. In the method, the angular thickness $\Delta b$ of the HI disk is measured along an annulus ring of radius $R$, which is related to  $R$ and $l$ by 
\begin{equation}
\Delta b = {\rm arctan} \left({ z_0 \over {R_0 \cos l + \sqrt{R^2 - R_0^2 \sin^2 l}}}\right).
\end{equation}

The longitudinal variation of $\Delta b$ is uniquely related to the galacto-centric distance $R$, and is as a function of $V(R)$. Since this method measures the averaged kinematics of the Galaxy, it represents a more global rotation curve than those based on the previous described methods. 

\section{Rotation in the Galactic Center} 

For its vicinity the Galactic Center rotation has been studied in most detail among the spiral galaxies, particularly in the nuclear region around Sgr A$^*$ nesting the super massive black hole. 
     
\begin{figure}
\begin{center}
\includegraphics[width=7.5cm]{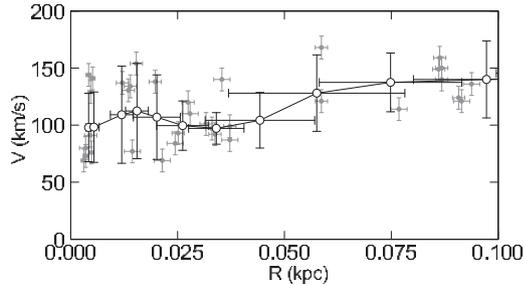}    
\end{center}
\caption{Rotation velocities in the Galactic Center obtained by terminal-velocity method using PV diagrams (grey dots: measured values; circles: running averaged values)(Sofue 2013).   
} 
\label{GCrc}
\end{figure}

Extensive CO-line observations of the Galactic Center have provided us with high quality and high-resolution longitude-velocity (LV) diagrams in the Galactic Center (e.g., Dame et al. 2001; Oka eta l. 1998). Figure \ref{GCrc} shows the rotation curve in the central $\sim 100$ pc as obtained by applying the terminal-velocity method to these LV diagrams (Sofue 2013b).  

Further inside $\sim 0.1$ pc of the Galactic Center, infrared proper-motion measurements of circum-nuclear giant stars have shown that the stars are moving around Sgr A$^*$ in closed elliptical orbits. Analyses of the orbits showed high velocities that increase toward the nucleus, giving firm evidence for a massive black hole locating at the same position as Sgr A$^*$ of $\sim 4$ million solar masses (see section \ref{massiveblackhole}).

\section{Logarithmic Rotation Curve: From Black Hole to Dark Halo} 

A rotation curve covering wider regions of the Galaxy has been obtained by compiling the existing data by re-scaling the distances and velocities to the common galactic constants $(R_0, V_0)$=(8.0 kpc, 200 \kms)  (Sofue et al. 2009). 
Rotation velocities in the outermost Galaxy and beyond the disk were estimated by analyzing radial velocities of globular clusters and satellite galaxies (Sofue 2013b).

\begin{figure}
\begin{center}
\includegraphics[width=7.5cm]{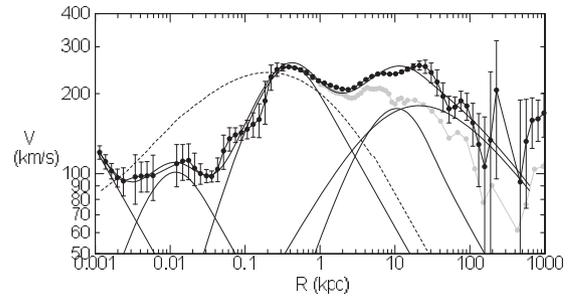}    
\end{center}
\caption{Logarithmic rotation curve of the Galaxy for $V_0=238$ \kms  and deconvolution into a central black hole, two exponential-spherical bulges, exponential flat disk, and NFW dark halo by the least $\chi^2$ fitting shown by thin lines. A classical de Vaucouleurs bulge shown by the dashed line is significantly displaced from the observation.   }
\label{logrc_mw}
\end{figure}

Figure \ref{logrc_mw} shows a logarithmic rotation curve of the entire Milky Way. The enlarged scale toward the center is useful to analyze the nuclear dynamics. The curve is drawn to connect the central rotation curve smoothly to the Keplerian motion representing the central massive black hole of mass $3.6 \times 10^6 \Msun$ (Ghez et al. 2005; Gillesen et al. 2009). This figure shows the continuous variation of rotation velocity from the Galactic Center to the dark halo. 

Detailed analyses of the rotation curve and deconvolution into mass components will be described in { chapter 4}.

\section{Uncertainty and Limitation of Rotation Curve Analyses} 

\subsection{Accuracy diagram}
The accuracy of the obtained rotation curve of the Milky Way is determined not only by the measurement errors, but also by the employed methods and the location of the observed objects in the Galactic disk (Sofue 2011). 

The most common method to calculate rotation velocity $V$ using the radial velocity $v_{\rm r}$ and distance $r$ yields 
\be
\Vrotr= {R \over R_0} \left({v_{\rm r} \over \sinl} +V_0 \right).
\label{eq-V}
\ee
The error of the derived velocity is given by
\be
\Delta \Vrotr=\sqrt{\delta V_{\rm vr}^2+\delta V_{\rm r}^2},
\ee
where
\be
\delta V_{\rm vr}={\partial V \over \partial v_{\rm r}} \delta v_{\rm r}, ~~~~
\delta {V_{\rm r}}={\partial V \over \partial r} \delta r.
\ee
Remembering equations (\ref{Rrl}) and (\ref{vfromvr}) we obtain
\be
\Delta \Vrotr 
=\left[{\left(R \over R_0 \sinl \right)^2\delta v_{\rm r}^2
+ \left(s ~V \over R^2 \right)^2 \delta r^2}\right]^{1/2}.
\label{eq-dvrad}
\ee 

The rotation velocity using the proper motion $v_{\rm p}$ is given by
\be
\Vrotp = - {R \over s} (v_{\rm p} + V_0 \cosl).
\label{eq-vrotprop}
\ee 
Knowing $R^2-s^2=R_0^2{\rm sin}^2 l$, we have the errors in this quantity as
\be
\Delta \Vrotp = 
{R \over s }  \left[ \delta v_{\rm p}^2 +  \left( R_0^2  v_{\rm p} {\rm sin}^2 l \over s R^2   \right)^2 \delta r ^2 \right]^{1/2}.
\label{eq-dvprop}
\ee 

\begin{figure} 
\begin{center}
\includegraphics[width=6cm]{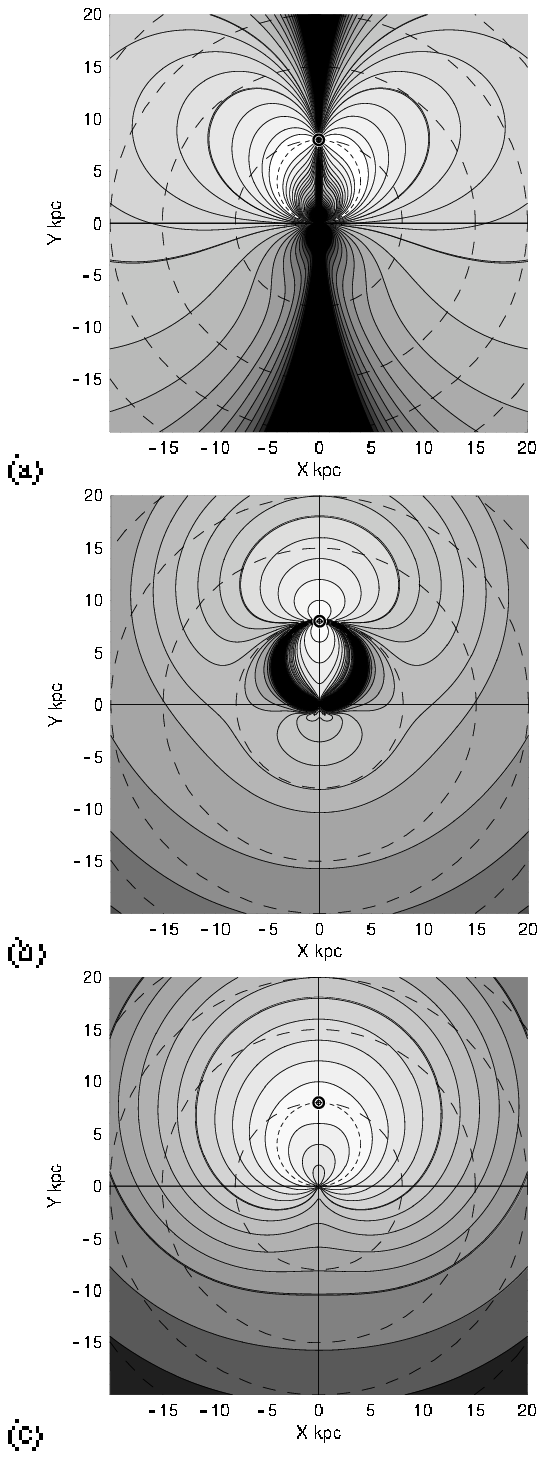}     
\end{center}
\caption{(a) Accuracy diagrams $\Delta \Vrotr(X,Y)$ for $\delta \vr=1$ \kms
 and 2\% distance error, (b) $\Delta \Vrotp(X,Y)$ for $\delta \mu=0.2$ mas y$^{-1}$, and (c) $\Delta \Vrotv(X,Y)$ for $\delta \vp /r=1 $ \kms kpc$^{-1}$ corresponding to $\delta \mu=0.21$ mas y$^{-1}$, $\delta \vr =1$ \kms, and 2\% distance error (Sofue 2011).}
\label{accu} 
\end{figure}

In figure \ref{accu} we present examples of the accuracy diagram of the radial velocity method for a set of measurement errors, $\delta \vr=1$ \kms and $\delta r/r=0.02 $, or 2\%. It is readily seen that the accuracy is highest along the tangent point circle (white regions), proving that tangent-point circle is a special region for accurate rotation curve determination. On the contrary, it yields the largest error near the singularity line running across the Sun and the Galactic Center, where the circular rotation is perpendicular to the line-of-sight.  

In figure \ref{accu}(b) we show an accuracy diagram of the proper-motion method for $\delta v_{\rm p}/r= 1~ {\rm km ~s^{-1} ~kpc^{-1}}$, or $\delta \mu=0.21 $mas y$^{-1}$, and  $\delta r/r= 0.02$. Contrary to the radial velocity method, the error becomes smallest along the Sun-GC line, but it is largest near the tangent point circle, where equation \ref{eq-dvprop} diverges. Thus, the tangent-point circle is a singular region in this method. 
This behavior is just in opposite sense to the case for the radial-velocity method, and the radial-velocity and the proper-motion methods are complimentary to each other.

 Figure \ref{accu}(c) shows the same, but for errors in velocity vector method using the 3D trigonometric measurement. This method yields much milder error variations in the entire Galaxy, showing no singular regions. 
   
\subsection{Implication} 
 The accuracy diagram, $\Delta \Vrotr(X,Y)$, demonstrates the reason why the rotation curve is nicely determined along the tangent-point circle, which is a special region yielding the highest accuracy determination of rotation velocity. The diagram also suggests that the butterfly areas at $l\sim 100-135\deg$ and $l\sim 225-280\deg$ are suitable regions for outer rotation curve work using the radial-velocity method. 

In the proper motion method, the most accurate measurement is obtained along the Sun-Galactic Center line, as was indeed realized by Honma et al. (2007). It must be also emphasized that the minimum error area is widely spread over $l \sim 120 - 250\deg$ in the anti-center region, as well as in the central region inside the tangent-point circle. On the other hand, the largest error occurs along the tangent-point circle, which is the singularity region in this method.

\section{Radial-Velocity and Proper-Motion Fields}

\begin{figure}
\begin{center}
\includegraphics[width=5cm]{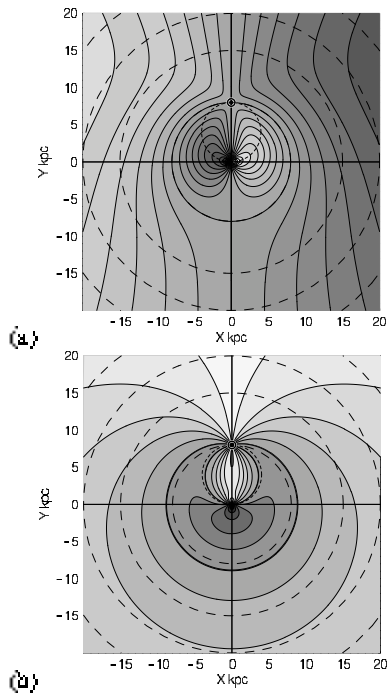}   
\end{center}
\caption{ 
(a) Radial velocity field $\vr(X,Y)$ with contour increment 20 \kms, and (b) proper motion field $\mu(X,Y)$ with contour increment 1 mas y$^{-1}$. The circle near the solar orbit is for 5 mas y$^{-1}$.  }
\label{fig-vfi} 
\end{figure}

If the rotation curve is determined, we are able to measure kinematical distances of any objects in the galactic disk by applying the velocity-space transformation, assuming circular rotation of the objects (Oort et al. 1958; Nakanishi and Sofue 2003, 2006, 2015a). The kinematical distance is obtained either from radial velocity or from proper motion using equations (\ref{eq-vr}) and (\ref{eq-vp}), and the velocities can be represented as a map on the galactic plane, as shown in figures \ref{fig-vfi}.

It may be worth to mention that the solar circle is a special ring, where the radial velocity is zero, $\vr=0$, whereas the proper motion $\mu$ has a constant value with  
\be
\mu_\odot=-{V_0 \over R_0} =\Omega_0,
\ee
which yields $\mu_{\rm \odot}=-5.26$ mas y$^{-1}$ for $R_0=8$ kpc and $V_0=$ 200 \kms.

VLBI trigonometric measurements have made it possible to apply this method to determine kinematical distance $r_\mu$. However, the radial-velocity field is currently used more commonly to map the distributions of stars and interstellar gas in the galactic plane because of its convenience. 

\begin{figure} 
\begin{center}   
\includegraphics[width=7cm]{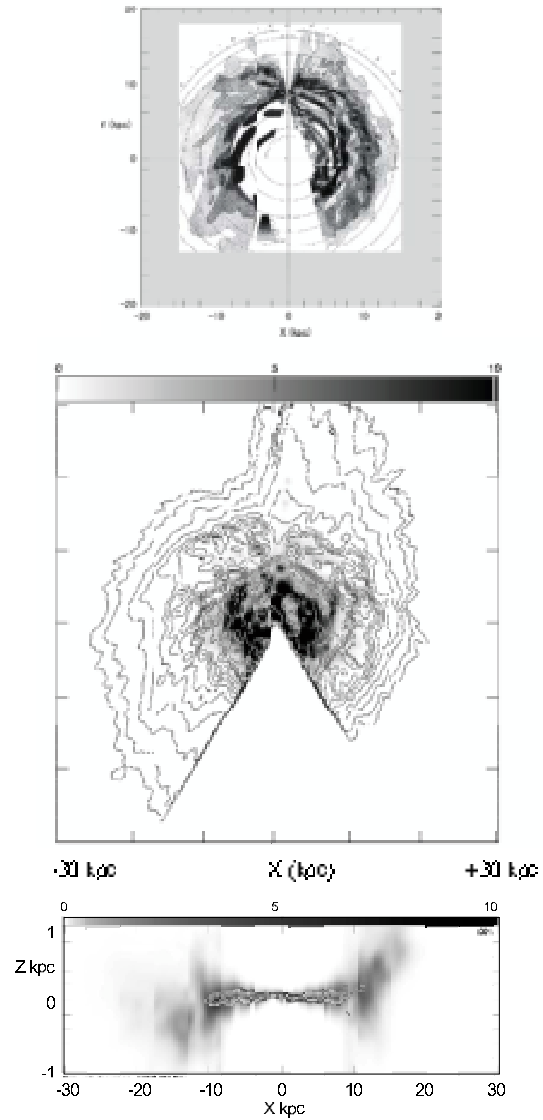} 
\end{center}
\caption{[Top] 2D HI map of the Galaxy, showing the distribution of volume density of the HI gas in the galactic plane on the assumption of circular rotation of the disk (Oort and Kerr 1958). [Middle] Surface densities of HI (contours) and H$_2$ (grey scale) gases obtained by integrating the 3D maps in the $Z$ direction (Nakanishi and Sofue 2016). [Bottom] HI and H$_2$ cross-section map of the Galaxy in the $(X,Z)$ plane across the Galactic Center. Shown here are the volume densities HI (extended thick disk in gray scale) and H$_2$ (thin disk by contours) (Sofue and Nakanishi 2016). Note, the $Z$ axis is enlarged by a factor of 4.
}
\label{HIH2face}
\end{figure}

\section{Velocity-to-Space Transformation}

If a rotation curve $V(R)$ is obtained or assumed, the radial velocity $\vr$ of an object in the galactic disk is uniquely calculated by equation (\ref{eq-vr}).  Figure \ref{fig-vfi} shows a thus calculated distribution of the radial velocity, or the velocity field.
Equation (\ref{eq-vr}) includes the distance and longitude. Therefore, the equation can be solved inversely to determine the position of an object using its radial velocity. This procedure is called the velocity-to-space (VTS) transformation.

By this method, the position of an object outside the solar circle is uniquely determined, but it has two-fold solutions for objects inside the solar circle. In order to solve this two-folding distance problem, further information such as apparent sizes of individual clouds and/or disk thickness is employed.

Applying the VTS transformation, galactic maps of HI and H$_2$ gases are obtained as follows. Column densities of HI and H$_2$ gases are related to the HI and CO line intensities as
\be
 N_{\rm HI}~[{\rm H~cm^{-2}}] = C_{\rm HI} \int T_{\rm HI}(v) dv~{\rm[K~km~s^{-1}]}
\label{eq-NHI} 
\ee
and
\be
 N_{\rm H_2}~[{\rm H_2~cm^{-2}}] = C_{\rm H_2} \int T_{\rm CO}(v) dv~{\rm[K~km~s^{-1}]},
\label{eq-NHtwho} 
\ee
where $T_{\rm HI}$ and $T_{\rm CO}$ are the HI and CO line  brightness temperatures, and $C_i$ are the conversion factors, and $C_{\rm HI}=1.82\times10^{18}$ [H cm$^{-2}$] and $C_{\rm H_2}\sim 2\times10^{20} f(R, Z)$ [H$_2$ cm$^{-2}$] are the conversion factors, and $f$ is a correction factor depending on $R$ (e.g. Arimoto et al. 1996).  

Volume density of the gas corresponding to radial velocity $\vr$ is obtained by
\be
 n_{\rm HI} = {dN_{\rm HI} \over dr}={dN_{\rm HI} \over d\vr}{d\vr \over dr}
 =C_{\rm HI} T_{\rm HI}(\vr){d\vr \over dr}
 \label{eq-denHI}
 \ee 
 and
\be
 n_{\rm H_2} = {dN_{\rm H_2} \over dr}={dN_{\rm H_2} \over d\vr}{d\vr \over dr}
 =C_{\rm H_2} T_{\rm CO}(\vr) {d\vr \over dr}.
 \label{eq-denH2}
 \ee 
These formulae enables us to create a face-on view of the ISM in the Galaxy, where the three-dimensional distribution of the volume densities of the HI and H$_2$ gases can be mapped. Fig. \ref{HIH2face} shows the thus obtained face-on view of the Galactic disk (Nakanishi and Sofue 2003, 2006, 2015).  

  		\chapter{ROTATION CURVES OF SPIRAL GALAXIES}
                
\section{Progress in Rotation Curve Studies Evidencing for the Dark Halo}

\begin{figure}
\begin{center}
\includegraphics[width=6cm]{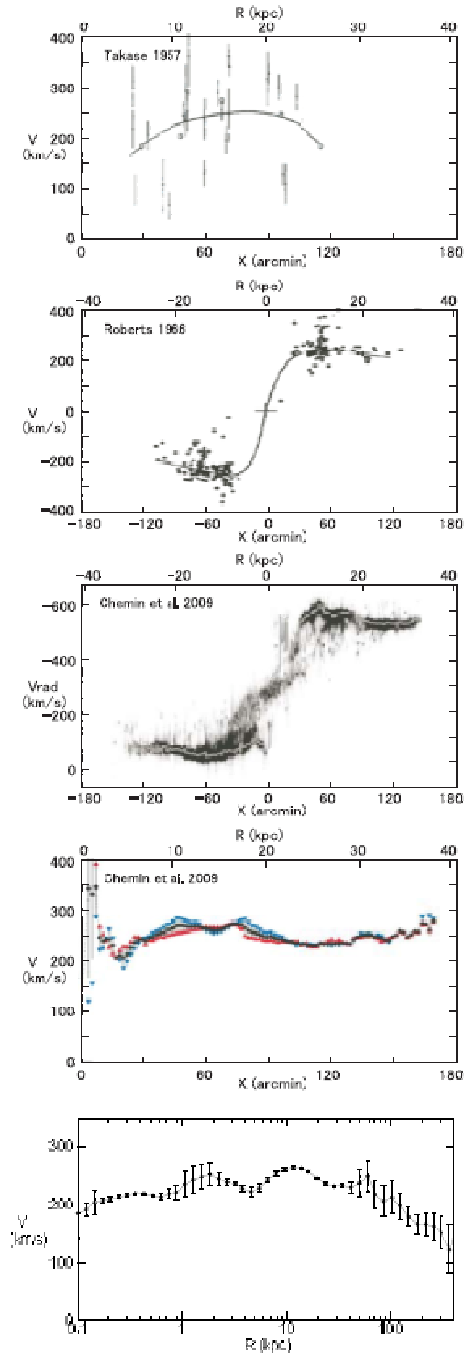}   
\end{center}
\caption{Half a century progress in the rotation curve of M31. From top to bottom: 1950's rotation curve using data by M.U. Mayall et al. of 1940's (Takase 1957); 
1960's in the HI 21-cm line (Roberts 1966); a modern PV diagram and rotation curve in HI, showing flat rotation up to 40 kpc (Chemin et al. 2009); and
a grand RC from the center to $\sim 300$ kpc in semi logarithmic scaling (Sofue 2013b).}
\label{progressm31} 
\end{figure}   
 
The rotation of galaxies was discovered a century ago when inclined spectra were observed across the nuclei of nearby galaxies. The modern era of rotation curves started in the 1950's when red-sensitive photographic plates were used to observe the H$\alpha$ $\lambda$6563 and [NII] $\lambda$6584 emission lines arising from HII regions. History review of developments in the galaxy rotation has been given in Sofue and Rubin (2001).  

Since the early measurements, flat rotation curves were routinely observed in spiral galaxies.  Figure \ref{progressm31} shows a progress in the rotation curve obtained for the spiral galaxy M31. Late in the last century, larger and advanced telescopes and detectors in optical, infrared, and radio observations combined with higher spectral resolution have allowed us to obtain higher accuracy rotation curves at farther distances. The generally observed 'flat rotation' in disk galaxies has shown the existence of massive dark halo around galaxies.
  
\section{Observations of Rotation Velocities} 

In this subsection we review the progress in obtaining rotation curves and related techniques and analysis methods. The content is based on our earlier review in Sofue and Rubin (2001), while recent topics and particular progress are added. In table \ref{rccat} we list the major papers in which rotation curve data are available in machine-readable formats, or in figures and tables. 

\def\ha{H$\alpha$}
\begin{table*}
\caption{Large catalogues of rotation curves in the two decades}  
\begin{tabular}{lllll}
\hline
Authors$^\ddagger$ (year) & Objects&Distances & Method &Catalogue type$^\dagger$ \\ 
\hline 
\\
Mathewson et al. (1992)&965 southern spirals&$<\sim 100$ Mpc& \ha/HI& RC/TF \\
Amram et al. (1992) & 21 NGC/UGC & Cluster & \ha & RC/VF/tab.\\
Makarov et al. (1997, 2001) & 135 edge-on & $\sim 100$ Mpc & \ha&RC\\
Vogt et al. (2004a,b)&329 spirals&$z<0.045$&\ha/HI, RC, M/L, FuP \\
Fridman et al. (2005) & 15 Sb/Sc/NGC& 10 - 70 Mpc &\ha/FP& RC/PV/VF\\
{\it GHASP} (2002-05) & 85 spirals &&\ha/FP& RC/VF \\
M{\'a}rquez et al. (2002) & 111 spipral/NGC & & \ha/HII &RC\\
Blais-Ouellette et al. (2004) & 6 Sb/Sc & $<\sim 20$ Mpc &FP& RC/VF\\
{\it URC} (1996-2007)& Spirals & Nearby&Av. of compil. &  Universal RC\\ 
Big FP H$_\alpha$ Survey (Hernandez, et al. 2005; & 21 barred spirals &
            2-38 Mpc & H$_\alpha$ FP & RC, Vf\\
~~~  Daigle et al. 2006; Dicaire et al. 2008) \\
Noordermeer et al. (2007) &19 S0/Sa/U,NGC &15 - 65 Mpc &\ha/HI&RC/PV/VF\\ 
{\it THINGS} (2008) &19 Nearby NGC & Nearby& HI &RC/PV/VF \\
Spano et al. (2008) & 36 NGC & Nearby & HI& RC\\ 
{\it DiskMass} (2010-13) & 146 face-on & $B<14.7$ & \ha/[OIII]/CaII/IS & RC/VF \\ 
Sofue et al. (1996); Sofue(2003, 2016) &$\sim 100$ Sb/Sc/NGC &Nearby+Virgo& \ha/CO/HI & RC/PV \\ 
\\ \hline\\
McGaugh et al. (2001)&36 LSB&&\ha/slit & RC \\
de Blok and Bosma (2002) &26 LSB/UGC & 3 - 45 Mpc&\ha/HI &RC/PV \\
Swaters et al. (2009) & 62 LSB dw/Ir/UGC &Nearby  & \ha & RC/PV\\
\\
\hline
\end{tabular} \\
 $^\dagger$ RC=rotation curve, VF=velocity field, PV=position-velocity diagram, FP=Fabry Perrot, FuP=fundamental plane; IS=integral fiber spectroscopy.
$^\ddagger$ {\it URC}: Persic et al. (1996), Salucci et al. (2007);{\it THINGS}: de Blok et al.  (2008); {\it GHASP}: Garrido et al. (2002-05); {\it DiskMass}: Bershady et al.(2010a,b), Martinsson et al. (2013a,b).
\label{rccat}
\end{table*}

\subsection{Optical observations}
 
Observations of optical emission lines such as \ha and [NII] lines sample population I objects, particularly HII regions associated with star forming regions in the galactic disk. These objects have small velocity dispersion compared to the rotation velocity, which allows us to derive circular velocities without suffering from relatively high velocity components by polulation II stars. The traditional method to derive rotation curves is long slit spectral observation along the major axis of disk galaxies (Rubin et al. 1982, 1985; Mathewson et al. 1992;  Amram et al.   1994; Corradi et al. 1991; Courteau 1997; 
 Sofue et al. 1998).

On the other hand, absorption lines, showing high velocity dispersion and slower rotation, manifest kinematics of population II stars composing spheroidal components and thick disk. Their line width is used to estimate the Virial mass through the pressure term in the equation of motion. In this paper, however, we concentrate on low velocity dispersion component for rotation velocities.

\begin{figure}
\begin{center}
\includegraphics[width=7cm]{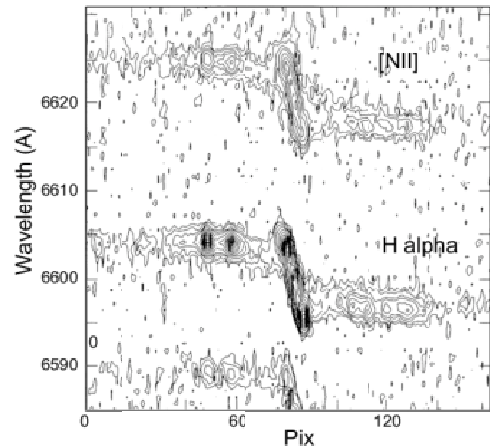}
\end{center}
\caption{Slit spectrum of the H$\alpha$ 6563A and [NII] 6584A lines along the major axis of Sb galaxy NGC 4527 (Sofue et al. 1999b)}
\label{slit}
\end{figure}

\subsection{Two-dimensional spectroscopy for velocity fields}

{ Fabry-Perot spectrographs are used to measure two-dimensional velocity fields in disk galaxies. Velocity fields include information not only of the global galactic rotation, but also non-circular stream motions due to spiral arms and bars (Vaughan 1989; Vogel et al. 1993;  Regan and Vogel 1994; Weiner  and  Williams 1996; Garrido et al. 2002, 2004; Kamphius et al. 2000; Vogel et al. 1993;  Hernandez, et al. 2005; Daigle et al. 2006; Dicaire et al. 2008; Shetty et al. 2007). }

The recent "DiskMass" survey of disk galaxies have observed largest sample of nearly face-on galaxies from the Upsala Galaxies Catalogue (UGC) brighter than $B=14.7$ with disk sizes 10 to 20 arcsec, and obtained their two-dimensional dynamical data (Bershady et al. 2010a,b; Martinsson et al. 2013a, 2013b; Westfall et al. 2014). They employed integral-field spectroscopy fiber instruments to measure stellar and ionized gas kinematics, covering [OIII]$\lambda$5007 and \ha lines. They also observed MgIb and CaII near-infrared triplet in stellar absorption for kinematics of spheroidal and old disk components.

\subsection{Infrared spectroscopy}
Spectral observations in the infrared wavelengths, including the Pa$\alpha$ and/or [Si VI] lines, and integral field analysis technique are powerful tools to reveal kinematics of dusty disks (Krabbe et al. 1997; Tecza et al. 2000). They are particularly useful in the nuclear regions of spiral galaxies, where dust extinction is significant, so that even such a red line as H$_\alpha$ is heavily obscured. Infrared observations are also powerful for mergers with dust buried nuclei. 

Rotation of outer disks of galaxies is crucial to derive the mass in the dark halo, while observations are still scarce because of the faint and small numbers of emission regions. Not only bright HII regions, but also planetary nebulae and satellite galaxies are used as test particles for determining the mass distribution in outer regions.

\subsection{HI and CO lines}

The 21-cm HI line is powerful to obtain kinematics of entire spiral galaxy, because the radial extent  in HI gas is usually much greater than that of the visible disk. HI measurements have played a fundamental role in establishing the flatness of rotation curves in spiral  galaxies (e.g., Bosma 1981a, b). 
   
The CO molecular lines in the millimeter wave range  are valuable in studying kinematics of the inner disk and central regions of spiral galaxies for their higher concentration toward the center than HI, which is deficient in the center, and for their extinction free nature against the central dusty disks (Sofue 1996, 1997).

\section{Methods to Determine Rotation Velocities}

A rotation curve of a galaxy is defined as the trace of terminal velocity on the major axis corrected for the inclination of the galaxy's disk. The observed lines are integration various velocity components along the line of sight through the galaxy disk. Hence, the intensity peaks do not necessarily represent the terminal velocities.

\subsection{Peak-intensity and intensity-weighted velocities}

The velocity at which the line intensity attains its maximum, which is called the peak-intensity velocity, is often adopted to represent the rotation velocity, method (Mathewson et al. 1992, 1996). 
More popular method is to measure the intensity-weighted velocities, which is sometimes approximated by a centroid velocity of half-maximum values of a line profile (Rubin et al.  1982, 1985). The intensity-weighted velocity is defined by
\be
 V_{\rm int}={\int I(\vr) \vr d\vr \over \int I(\vr) d\vr}.
\ee
Here, $I(\vr)$ is the intensity as a function of the radial velocity.
Rotation velocity is then given by
\be
\Vrot=(V_{\rm int}- V_{\rm sys})/{\rm sin}~i,
\ee
where $i$ is the inclination angle and
$V_{\rm sys}$ is the systemic velocity of the galaxy.  
 
However, the intensity-weighted velocity gives always underestimated rotation velocity because of the finite resolution of observation, due to which the plus and minus velocities in both sides of the nucleus are not resolved, but are averaged. Thus, derived rotation curves often start from zero velocity at the center. However, the nucleus is the place nesting a massive black hole and high density bulge core, where gas and stars are most violently moving. { Therefore, one must carefully apply the peak-intensity and/or intensity weighted velocity methods, because the methods often largely underestimate the central rotation velocities. However, they usually result in good approximation in sufficiently resolved outer disks. }

\subsection{Terminal-velocity and envelope-tracing method}

The terminal velocity  $V_{\rm t}$ in a position-velocity (PV) diagram is defined by a velocity, at which the intensity becomes equal to
\be
I_{\rm t}=[(\eta I_{\rm max})^2+I_{\rm lc}^2]^{1/2},
\ee
where $I_{\rm max}$ is the maximum intensity, and $I_{\rm lc}$ is the intensity representing the lowest contour, which is often taken at $\sim 3$ rms noise level in the PV diagram. The fraction $\eta ~(0.2 \sim 0.5$)  represents the critical intensity level of the line profile.

The rotation velocity is then derived as
\be
\Vrot=(V_{\rm t}-V_{\rm sys}) / {\rm sin}~i~
-(\sigma_{\rm obs}^2 + \sigma_{\rm ISM}^2)^{1/2},
\ee 
where $\sigma_{\rm ISM}$ and $\sigma_{\rm obs}$ are the velocity dispersion of the interstellar gas and the velocity resolution of observations, respectively.
The interstellar velocity dispersion is of the order of $\sigma_{\rm ISM} \sim 7$ to 10 \kms, while $\sigma_{\rm obs}$ depends on instruments.

\subsection{Iteration methods using position-velocity (PV) diagram and 3D cube}
A reliable method to estimate the rotation velocity is the iteration method, with which observed PV diagram is reproduced by the finally obtained rotation curve (Takamiya and Sofue 2000). In this method, an initial rotation curve RC$_0$ settled from the observed PV diagram using by any method as above. Using RC$_0$ and the observed radial distribution of the intensity, a PV diagram, PV$_1$, is constructed. The difference between PV$_1$ and PV$_0$ is then used to correct RC$_0$ to obtain a corrected rotation curve, RC$_1$. This RC$_1$ is used to calculate another PV diagram PV$_2$ using the observed intensity distribution, from which the next iterated rotation curve, RC$_2$ is obtained by correcting for the difference between PV$_2$ and PV$_0$. This procedure is iteratively repeated until PV$_i$ and PV$_0$ becomes identical within the error, and the final PV$i$ gives the most reliable rotation curve.

\begin{figure}
\begin{center}
\includegraphics[width=6cm]{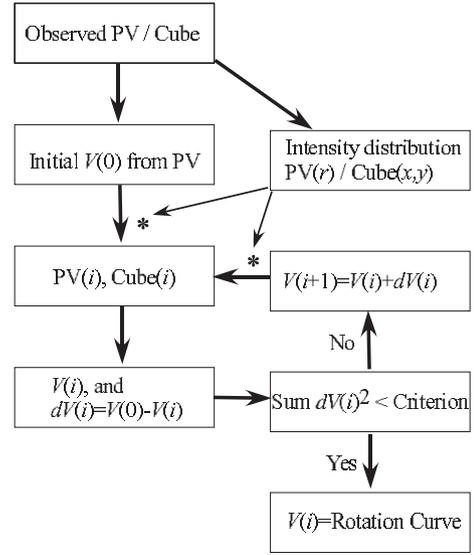}  
\end{center}
\caption{Iteration method of rotation curve fitting using PV diagrams and/or 3D cube.}
\label{fig_RCIdia}   
\end{figure}

\begin{figure}
\begin{center}
\includegraphics[width=6cm]{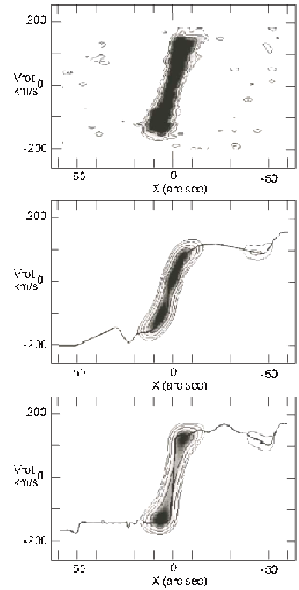}  \\ 
\end{center}
\caption{Iteration method: a PV diagram of NGC 4536 in the CO line (top panel),  an approximate rotation curve using the peak-intensity method  and corresponding PV diagram, the final rotation curve and reproduced PV diagram (bottom).}
\label{fig_RCIngc}   
\end{figure}

\subsection{3D cube iteration}
The PV diagram along the major axis represents only one dimensional velocity information along the major axis. The next-generation measurement of rotation velocities would be to utilize three-dimensional spectral data cube of the entire galaxy. This will be particularly useful for optically thin radio lines in the central regions such as HI and CO lines. The reduction procedure would be similar to that for the PV iteration method. We first settle an approximate RC, and calculate a data cube using this RC and observed intensity distribution. The calculated cube is then compared with the original cube, and the difference is used to correct for the initially assumed rotation curve. This procedure is repeated iterative to minimize the difference between the calculated and observed cubes. In 3D iteration method, the entire data cube is analyzed without losing information, and we would be able to reach an ultimate rotation curve of a galaxy. The method will be particularly useful for obtaining a rotation curve in the central region, where observational resolution is not sufficient.

\subsection{Tilted-ring method for rotation velocity} 

The rotation velocity $\Vrot$, inclination angle $i$,  and observed radial velocity $\vr$ relative to the systemic velocity are related to each other as
\begin{equation}
\vr(r,\theta) = V_{\rm rot}(r) \cos \theta \sin i,
\end{equation}
where $\theta$ is azimuth angle in the disk of a measured point from the major axis. Obviously the rotation velocity and inclination are coupled to yield the same value of $\vr$. 

The most convenient way to derive a rotation curve is to measure radial velocities along the major axis using position-velocity diagrams, as described in the previous  subsection. Thereby, inclination angle $i$ has to be measured independently or assumed.  Given the inclination, the rotation velocity is obtained by
\begin{equation}
V_{\rm rot}={\vr \over  \sin i}
\label{eq_vrot}
\end{equation}
with $\vr=\vr(r,0)$. 

Inclination angle $i$ is measured from the ratio of major to minor axial lengths of optical isophotes of a galaxy. An alternative way is to compare the integrated HI line width with that expected from the Tully-Fisher relation (Shetty et al. 2007). However, as equation \ref{eq_vrot} shows, the error in resulting velocity is large for small $i$, and the result diverges for a face-on galaxy.  

If a velocity field is observed, coupling of the rotation velocity and inclination can be solved using the tilted-ring technique to determine either rotation velocity or the inclination (Bosma 1981; Begeman 1989; J{\'o}zsa et al. 2007). The radial velocity $\vr$ is related to position angle $\phi$ and azimuthal angle $\theta$ as
\begin{equation}
f(\theta, i)={\vr \over v_{\rm r, max}} ={\cos} ~\theta(\phi,i),  
\label{eq_tilt}
\end{equation}
with
\be
\theta(\phi,i)={\rm atan} \left({{\rm tan}~ \phi \over {\cos}~i} \right),
\ee
and  $v_{\rm r,max}$ is the maximum value along an annulus ring.

\begin{figure}
\begin{center} 
\includegraphics[width=7cm]{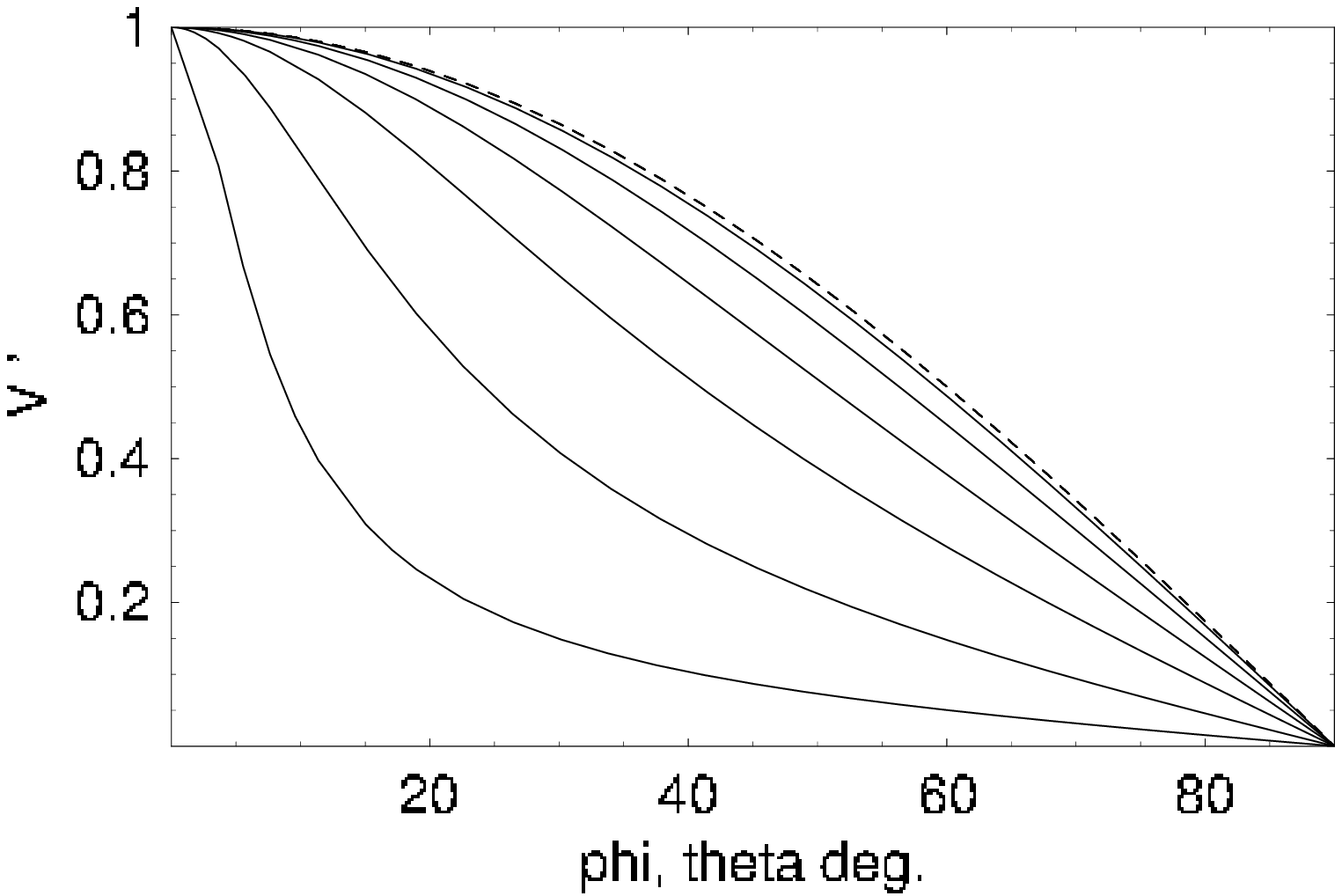}    
\end{center}
\caption{Variation of $V'=\vr / v_{\rm r, max}$ along a tilted-ring as a function of position angle $\phi$ (full line) for different inclinations (from bottom lines: $i=85\deg, ~75\deg, ~60\deg, ~45\deg, ~30\deg$ and $15\deg$).  Variation against azimuthal angle $\theta$ is shown dashed. } 
\label{tiltedring}
\begin{center}  
\includegraphics[width=5cm]{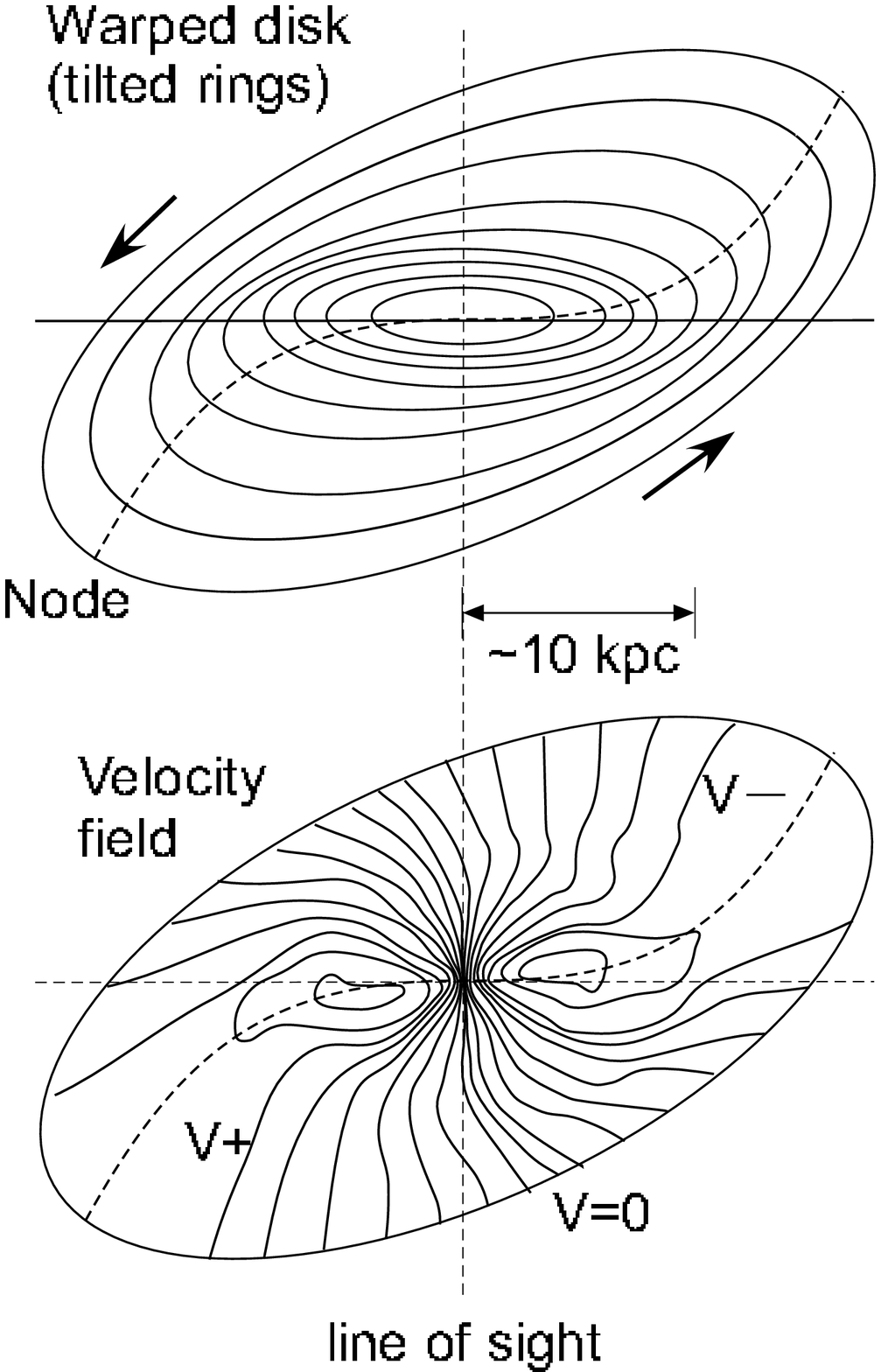}    
\end{center}
\caption{Tilted rings and velocity field fitted to HI velocity fields of NGC 5055 (Bosma 1981a).  } 
\label{tiltedringbosma}
\end{figure}

Figure \ref{tiltedring} shows variations of $\cos~ \theta(\phi)$, or $\vr$ normalized by its maximum value along an annulus ring, as functions of $\phi$ and $\theta$ for different values of inclination. The functional shape against $\phi$ is uniquely dependent on inclination $i$, which makes it possible to determine inclination angle by iterative fitting of $\vr$ by the function. Once $i$ is determined, $\Vrot$ is calculated using $\vr$. Thus, both the inclination and $\Vrot$ are obtained simultaneously. 

In often adopted method, the galactic disk is divided into many oval rings, whose position angles of major axis are determined by tracing the maximum saddle loci of the velocity field. Along each ring the angle $\theta$ is measured from the major axis. Observed values of $f(\theta, i)$ are compared with calculated values, and the value of $i$ is adjusted until $chi^2$ gets minimal stable. The value of $i$ yielding the least $\chi^2$ is adopted as the inclination of this ring. This procedure is applied to neighboring rings iteratively to yield the least $\chi^2$ over all the rings. Begeman (1989) has extensively studied the tilted-ring method, and concluded that it is not possible to determine inclinations for galaxies whose inclination angles are less than $40\deg$. Di Teodoro and Fraternali (2015) recently developed a more sophisticated method incorporating the velocity field as a 3D cube data (velocity, $x$ and $y$) to fit to a disk rotation curve. 

\subsection{Tilted-ring method for inclination} 
The tilted-ring method is useful for highly and mildly inclined galaxies for velocity determination, but it fails when the galaxy is nearly face-on.
This property, in turn, may be used to determine the inclination by assuming the rotation velocity, remembering that the inclination is written as
$
\sin i= \vr / \Vrot,
$
which means that the inclination can be determined by measuring $\vr$, if $\Vrot$ is given. 

This equation is used for determination of inclination using the Tully-Fisher relation: The intrinsic line width is determined by the disk luminosity, and is compared with observed line width to estimate the inclination angle. The equation can be used to determine inclinations of individual annulus rings, if the rotation curve is given. This principle has been applied to analyzing warping of the outer HI disk of the face-on galaxy NGC 628 (Kamphuis and Briggs 1992) and M51 (figure \ref{m51warp}: Oikawa and Sofue 2014).

\section{Galaxy Types and Rotation Curves} 

\subsection{Observed rotation curves}

\begin{figure} 
\begin{center} 
\includegraphics[width=7cm ]{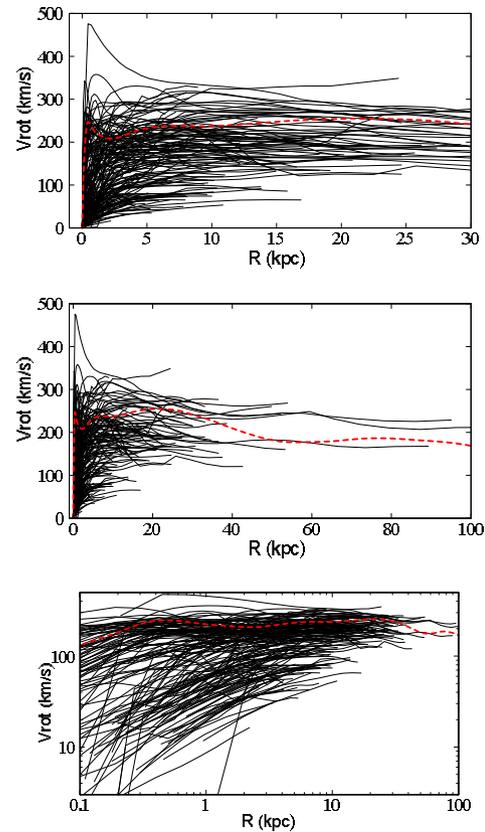}    
\end{center}
\caption{Rotation curve compilation (Sofue 2015), which includes those from Sofue et al. (1999, 1999; 2003, 1999);  Ryder et al. (1998);  Hlavacek-Larrondo et al. (2011a,b);  Erroz-Ferrer et al. (2012);  Gentile et al. (2015);   Olling R.~P. (1996);  Whitmore  \& Schweizer (1987);  Richards et al. (2015) ;Scarano et al. (2008);  Gentile et al. (2007);   Marquez et al. (2004); de Blok et al. (2008); Garrido et al. (2005); Noordermeer et al. (2007); Swaters et al. (2009); Martinsson et al. (2013); Bershady et al.(2010a,b).
} 
\label{allrc} 
\end{figure} 

\begin{figure}
\begin{center}
\hskip 1mm \includegraphics[width=7cm]{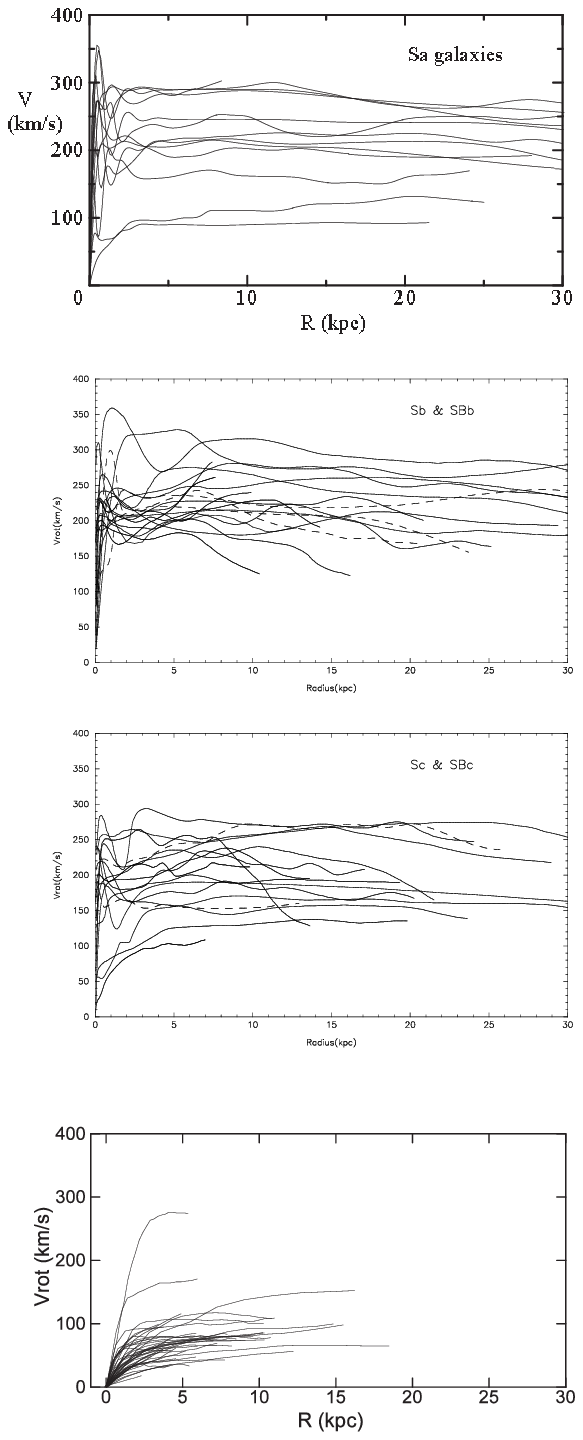}  
\end{center}
\caption{Rotation curves (RC) for Sa type galaxies (top: Noordermeer et al. 2007), Sb (full lines) and barred SBb (dashed lines) (second panel); Sc and SBc (third panel: Sofue, et al. 1999), and dwarf and LSB galaxies (bottom: Swaters et al. 2009)}
\label{rc-type}    
\end{figure}

Figure \ref{allrc} shows rotation curves published in the two decades as compiled from the literature by Sofue (2016), and figure \ref{rc-type} shows those for galaxies types from Sa to Sc. 
The shapes of disk and halo rotation curves are similar to each other for different morphologies from Sa to Sc, from less massive to massive galaxies. This suggests that the form of the gravitational potential in the disk and halo is rather universal over the galaxy types.
 
\begin{figure}
\begin{center}
\includegraphics[width=7cm]{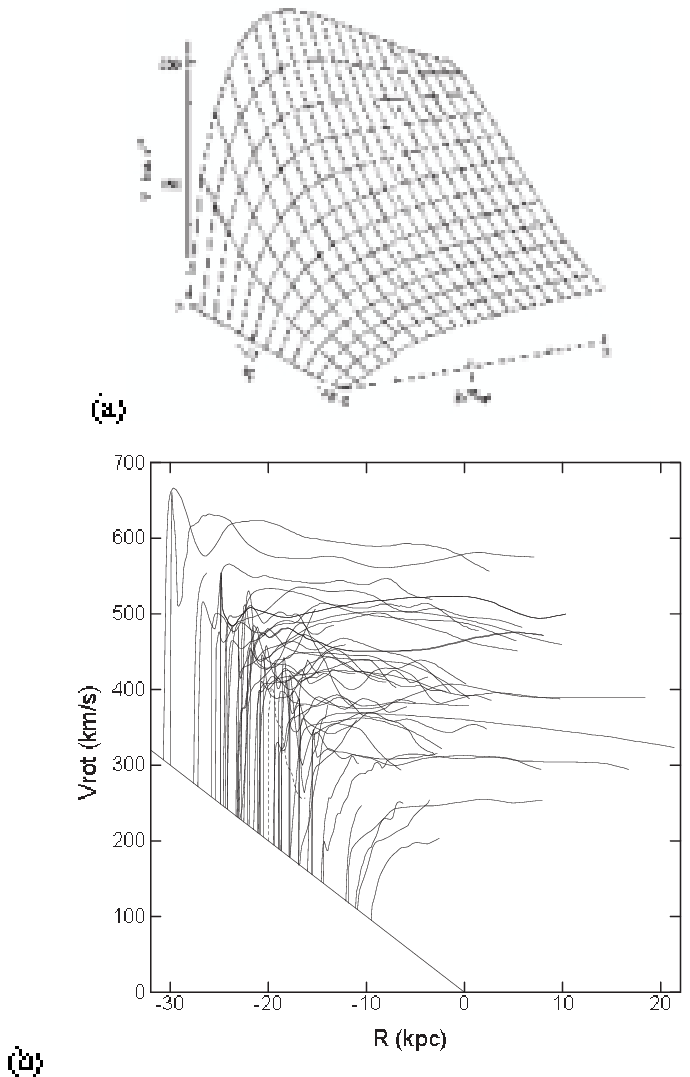}  
\end{center}
\caption{(a) Universal rotation curves obtained from 967 spiral galaxies by solid lines (Persic and Salucci 1996).  
(b) Observed rotation curves from figure \ref{allrc} with the origins shifted according to the disk rotation velocities (averages from $R=1$ to 10 kpc), approximately from Sa (top-left), Sb to Sc (bottom-right).} 
\label{universalrc}
\end{figure}

Figure \ref{allrc} shows examples of rotation curves of nearby spiral galaxies obtained by combining optical (mainly \ha) and radio (CO and HI) observations.
There is a marked similarity of form of rotation curves for galaxies with different morphologies from Sa to Sc (figure \ref{rc-type}). The forms may be classified into three groups: the centrally peaked, shoulder rise, and rigid-body rise types. The three types are observed mainly in massive and large-diameter galaxies, medium sized galaxies, and less massive Sc and dwarf galaxies, respectively.

\subsection{Universal rotation curve}
Massive Sa to Sb galaxies show higher central velocities than less massive Sc galaxies (figure \ref{universalrc}). On the contrary, dwarf galaxies show slower central rise in a rigid-body fashion. Massive galaxies have universally flat rotation, while less massive galaxies show monotonically increasing rotation curve (Persic et al. 1996). 
The observed rotation curves may be approximated by simple function (Persic et al. 1996; Courteau 1997; Roscoe 1999), which is dependent on the luminosity and radius of a galaxy.

\subsection{Flatness and similarity of rotation curves}

\begin{figure} 
\begin{center} 
\includegraphics[width=7cm]{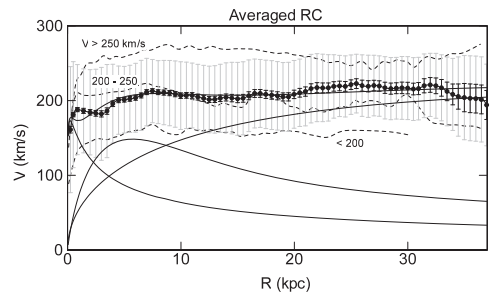}  
\end{center}
\caption{Black circles show Gaussian averaged rotation curve from all galaxies listed in Sofue et al. (1999). Long and short bars are standard deviations and standard errors, respectively. Thin lines show the least-$\chi^2$ fitting by de Vaucouleurs bulge (scale radius 0.57 kpc, mass $9.4\times 10^9\Msun$), exponential disk (2.7 kpc, $3.5\times 10^{10}\Msun$), and NFW dark halo (35 kpc, $\rho_0=3\times 10^{-3} \Msun{\rm pc}^{-3}$).  Three dashed lines are averaged rotation curves of galaxies with maximum velocities greater than 200 \kms, between 200 and 250 \kms, and below 200 \kms, respectively, from top to bottom. }
\label{rc_sum}  
\end{figure}

The flatness of the overall shape of entire rotation curves applies to any mass ranges of galaxies. Figure \ref{rc_sum} shows averaged rotation curves of galaxies from the sample of Sofue et al (1999), categorized into three groups by dashed lines, one those with maximum rotation velocity greater than 250 \kms, and the second between 200 and 250 \kms, and the third slower than 200 \kms. 

The thick line shows a Gaussian averaged rotation curve of all the sample galaxies, where long and short bars denote the standard deviation and standard error of the mean value, respectively, in each radius bin at 0.5 kpc interval with Gaussian averaging width of 0.5 kpc.

\section{Dark Halo: Grand Rotation Curves}

\begin{figure}
\begin{center}     
\includegraphics[width=7cm]{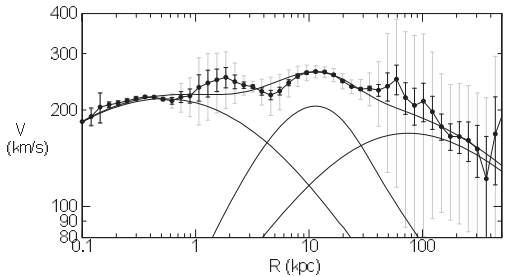}   
\end{center}
\caption{Logarithmic rotation curve of M31. Thin lines show the least-$\chi^2$ fit by the bulge, disk and dark halo components. } 
\label{GRCm31mw}  
\end{figure}  

The outermost rotation curve beyond the flat disk is still not well determined for external galaxies because of the lack in extended disk and weak emissions. In order to solve this difficulty, non-coplanar objects such as outer globular clusters and satellite galaxies have been often used to estimate the dark halo masses applying the Virial theorem.

Similarly to the Milky Way as described in section 2, orbiting satellite galaxies can be used to obtain expected circular velocities within several hundred kpc of the parent galaxy (Sofue 2013b). The technique has been applied to M31 to derive a grand rotation curve (GRC) within $\sim 300$ kpc by combining disk rotation velocities and radial velocities of satellite galaxies and globular clusters. Figure \ref{GRCm31mw} shows the logarithmic GRC and model fitting by bulge, disk and dark halo components (Sofue 2015a). 
{ The whole rotation curve up to $\sim 40$ kpc is well represented by the pseudo-isothermal halo model (Burkert 1995), predicting a flat rotation, while the outermost region beyond 50 kpc seems to be better represented by the Navaro-Frenk-White (1996, 1997) (NFW) model (see chapter 3). }

\section{Nuclear Rotation Curves and Black Holes} 

\subsection{Innermost velocities}

Central rotation curves have been produced for a number of galaxies by a systematic compilation of PV diagrams in the \halpha\ and CO lines (Rubin et al. 1997; Sofue et al. 1997, 1998, 1999; Bertola et al. 1998;). 
Many spirals exhibits rapid central velocities, suggesting massive compact nuclear objects (van der Marel et al. 1994; Kormendy  and  Richstone 1995; Richstone, et al. 1998, Bertola et al. 1998; Ferrarese 1999; Kormendy  and  Westpfahl 1989; Kormendy 2001). 

VLBI water maser observations of the spiral galaxy NGC 4258 showed a disk of radius 0.1 pc in Keplerian rotation, indicating the first firm evidence for a massive black hole of mass of $3.9 \times 10^{7}\Msun$ (Nakai et al. 1993; Watson  and  Wallim 1994; Miyoshi et al. 1995; Herrnstein et al. 1999). Further VLBI observations of the water maser line have revealed a rapidly rotating nuclear torus of sub parsec scales in several nearby active galactic nuclei (Haschick et al. 1990; Trotter et al. 1998; Sawada-Satoh et al. 2000; Greenhill et al. 1996). 

\begin{figure} 
\begin{center}
\includegraphics[width=7cm]{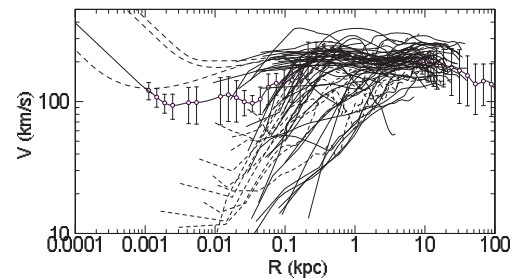}   
\end{center}
\caption{Logarithmic rotation curves of nearby spiral galaxies  with typical resolutions, $\sim 2-10''$ (thick lines) and those from Virgo CO survey ($\sim 1''$, thick dashed lines), compared with the Milky Way rotation curve (circles with error bars). Thin dashed lines show NGC 224, 1068 and 4258 as interpolated with their black holes by horizontal straight lines.}
\label{logrc_all}  
\end{figure} 

\subsection{Logarithmic presentation}

Logarithmic presentation of rotation curves is powerful to discuss the innermost dynamics in relation to the galactic structure. Figure \ref{logrc_all} shows rotation curves of nearby galaxies and Virgo samples in logarithmic presentation. Also shown are the rotation curves of nearby galaxies with known super massive black hole, where black holes and outer rotation curves are interpolated by dashed lines.  The thick line with dots and error bars are the rotation curve of the Milky Way (Sofue 2013b).  The figure demonstrates that, except for the Milky Way, the resolution is not sufficient, and higher resolution observations  such as using ALMA are a promising subject for the future for linking the central black hole to the galactic structure.  

\subsection{Effect of finite resolution}

Simulation in figure \ref{fig_RCIngc} reveals the effect of the finite resolution on the observed PV diagram as shown. Central rotation curves derived from observed PV diagrams generally give lower limits to the rotation velocities.  
Angular resolution and high dynamic range are crucial for the analysis of central kinematics (Rubin et al. 1997; Sofue et al. 1998, 1999a; Bertola et al. 1998). Also crucial is the interstellar extinction in optical observations of the nuclear dusty disks. To avoid this difficulty, radio line observations, particularly the CO lines, are powerful because of the negligible extinction and high concentration in the nuclei (Sofue et al. 1999a).
 
\section{Shapes of Rotation Curves and Galaxy Types}

Figure \ref{rc-type} shows observed rotation curves of Sa, Sb, Sc and low-surface brightness galaxies.
Individuality of observed rotation curves of many spiral galaxies can be explained by the difference in dynamical parameters of the bulge, disk and dark halo. Without any particular exception, the observed rotation curves can be reproduced by properly choosing the masses and scale radii of the mass components. 
 
\subsection{Sa to Sc galaxies} 

In each mass component of the bulge, disk and dark halo, it is known that the larger is the scale radius, the more massive is the component, and therefore the higher is the corresponding rotation velocity (Sofue 2016). This general characteristics applies to the comparison among different galaxies. Thus, the general tendency of differences of rotation curve shapes among the galaxy types can be simply attributed to their mass and size differences. Namely, earlier-type (Sa, Sb) galaxies are generally more massive and larger in size, and hence show higher rotation velocity, than later (Sc) type galaxies.

The most significant difference in the shape occurs in the central rotation curves: Sa and Sb galaxies, including the Milky Way, have high-velocity rotation near the nucleus, while Sc galaxies show slower and rigid-body like rotation in the center. This means that Sa and Sb galaxies have large and massive central bulge. On the other hand, Sc galaxies have smaller and less massive bulge. 

Difference, though not so significant, appears in the outermost rotation curve: Earlier galaxies show flat or slowly declining rotation in the outermost region, while later type galaxies show monotonically increasing rotation.  

\subsection{Barred galaxies}
Kinematics of barred galaxies is complicated due to the non-circular streaming motion superposed on the circular motion (e.g., Bosma 1981a,b).
Considering that a half or more of observed galaxies exhibit bars, it is not easy to discuss particular differences of rotation curves between bar and non-bar galaxies. Rotation curves for SBb and SBc galaxies shown by dashed lines in figure \ref{rc-type} are not particularly distinguishable from normal galaxies. 

Another effect of bars on rotation curves would be statistical underestimation of rotation velocity: Since interstellar gas is streaming along the bar (Hunter  and  Gottesman 1996; Buta et al.  1999; Kuno et al.  2000), observed velocities are close to rigid-body motion of the bar potential, and hence slower than circular velocity. 
 Considering the larger probability of side-on viewing of a bar than end-on probability, simply derived rotation velocity from the radial velocities is statistically underestimated. 

{ For barred galaxies, more sophisticated modeling of non-circular motions is inevitable based on 2D velocity measurements, such as the DISKFIT method proposed by Spekkens and Sellwood (2007) or the numerical method based on $N$-body simulations (Randriamampandry et al. 2015). In fact in the decade, an extensive 2D spectroscopy has been obtained for a large number of barred spiral galaxies by Fabry-Perot \ha observations (e.g., Dicaire et al. 2008), which will further open a new era of non-axisymmetric galactic dynamics, overcoming the difficulties as raised in table \ref{parameters}.  }

\subsection{Dwarf and low surface brightness (LSB) galaxies}

Within the decades, a large number of low surface brightness (LSB) galaxies have been found (Schombert  and  Bothun 1988; Schombert et al. 1992). Dwarf and LSB galaxies show slow rotation, monotonically rising until their galaxy edges (de Blok et al. 1996, 2001;  de Blok 2005; Swaters et al. 2000, 2001, 2009; Carignan  and  Freeman 1985;  Blais-Quellette et al. 2001; Noordermeer et al. 2009; figure \ref{swaters2009LSB}).  Blue compact galaxies also show that rotation curves rise monotonically to the edges of the galaxies (\"Ostlin et al. 1999).  Also, dwarfs with higher central light concentrations have more steeply rising rotation curves, similarly to spirals. 

Observations show that the mass-to-luminosity (M/L) ratio of dwarfs and LSB is usually higher than that normal spirals, indicating that they are more dark matter dominant than normal spirals (Carignan 1985; Jobin  and  Carignan 1990; Carignan  and  Freeman 1985; Carignan  and  Puche 1990a,b; Carignan  and  Beaulieu 1989; Puche et al. 1990, 1991a, b; Lake et al. 1990; Broeils 1992; Blais-Ouellette et al. 2001; Carignan et al. 2006).

\subsection{Interacting and irregular galaxies}

Rotation curves for irregular and interacting galaxies are not straightforward. Some irregular galaxies exhibit quite normal rotation curves, whereas some reveal apparently peculiar rotations. Hence, it is difficult to deduce a general law property to describe the curves, but individual cases may be studied in case by case. We may raise some examples below.

The Large Magellanic Cloud is the closest dwarf galaxy interacting with the companion Small Magellanic Clouds and the parent Milky Way. Despite of the strong gravitational interaction, the rotation curve is nearly flat at $\sim 100$ \kms (Kim et al. 1998). The dynamical center inferred from the kinematical center is displaced from the optical center of the bar center, suggesting a massive component that is not centered by a stellar bulge.  The dark bulge suggests an anomalously high M/L ratio in the dynamical center (Sofue 1999).

\begin{figure}
\begin{center} 
\includegraphics[width=7cm]{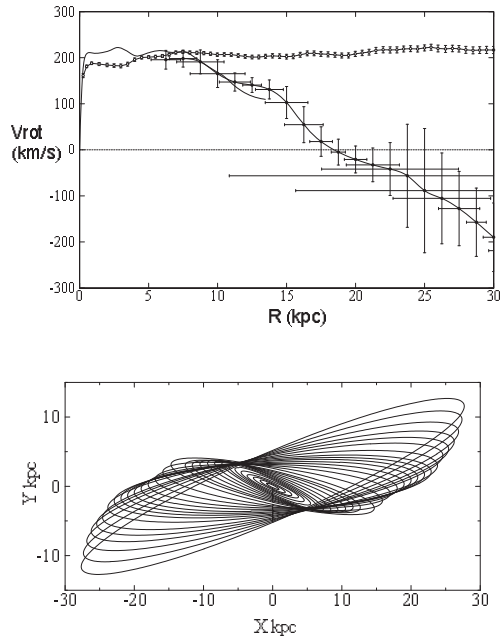}  
\end{center}
\caption{[Top] Anomalous rotation curve of M51 showing faster decrease than Keplerian beyond 8 kpc followed by apparent counter rotation, compared with averaged rotation curve of spiral galaxies. [Bottom] Warped disk calculated for rotation curve assuming a universal rotation. The line-of-sight is toward top. (Oikawa and Sofue 2014)}
\label{m51warp}
\end{figure}

The peculiar galaxy M51 (NGC 5194) interacting with the companion NGC 5195 exhibits an anomalous rotation curve, which declines more rapidly than  Keplerian in the outer disk (figure \ref{m51warp}). Even counter rotation is observed in the outermost HI ring (Rots et al. 1990; Appleton et al. 1986). Kinematics of M51 has been observed at various wavelengths, which all indicates the rotation anomaly. This apparent rotation anomaly is explained by the warping of galactic disk, assuming that the galaxy has intrinsically normal flat rotation curve (Oikawa and Sofue 2014). Figure \ref{m51warp} shows a calculated warping of the disk for a given normal rotation curve, where the disk is nearly flat and then bends at $r=7.5$ kpc steeply. 

The peculiar galaxy M82 is a companion to the giant spiral M81. It shows an exceptionally peculiar rotation property. It has a normal steep rise and high rotation near the center, but it exhibits a significantly declining rotation, obeying the Keplerian law beyond the peak. This can be well explained and modeled by tidal truncation of the outer disk and dark halo by the close encounter with the massive parent galaxy M81 (Sofue 1992).
   
\subsection{Activity and rotation curves}
 
Active galaxies like starburst galaxies, Seyferts, LINERs, and those with nuclear jets appear to show no particularly peculiar rotation. Even such an active galaxy like NGC 5128 (Cen A) shows a rotation curve similar to a normal galaxy (van Gorkom et al. 1990). 
The edge on galaxy NGC 3079 has strong nuclear activity and is associated bipolar lobes in radio continuum and ionized and high-temperature gases. However, its rotation properties are quite normal with very high central velocities (Sofue  and  Irwin 1992; Irwin  and  Sofue 1992).

While these galaxies show a steep central rise of rotation curve, it is not peculiar, but generally observed in normal massive galaxies without any pronounced activity. Thus, the global rotation and mass distribution of galaxies having activity are considered to be normal. This means that the nuclear activity is not directly produced by the fundamental dynamical structure, but is caused by more temporal dynamics and/or ISM phenomena in the central deep gravitational potential, such as a sudden inflow triggered by intermittent feeding of circum nuclear gas and stars.

\section{Statistical Properties of Rotation Curves: Tully-Fisher Relation}

Rubin et al. (1985) used their synthetic rotation curves to derive dynamical masses interior to the Holmberg radius ($R_{25}$) of spiral galaxies, and showed a clear correlation with the $H$-band infrared luminosity. Instead of the total dynamical mass, the maximum rotation velocity at a few galactic-disk scale radii was shown to be related to the luminosity, as observed as one-half the velocity width of an integrated 21 cm HI line velocity profile. The relation is called the Tully-Fisher relation (Tully and Fihser 1977; Aaronson  and  Mould 1986; Mathewson et al. 1992, 1996; Masters et al. 2008, 2014), and is one of the major tools to derive intrinsic luminosities of distant galaxies for measurement of the Hubble constant.

\section{High Redshift Rotation Curves}

Rotation and dynamics of high-redshift galaxies at cosmological distances are one of the major subjects in the new era of sensitive observations in the 21st century. Rotation curves of sub-$z$ ($z<1$) galaxies had been observed already in the decades (Simard  and  Prichet 1998; Kelson et al. 2000; Vogt et al. 1993, 1996, 1997). However, the modern era of higher redshift rotation curves was opened in this century. 

Erb et al. (2003) used Keck I, II telescopes and VST to obtain near-infrared slit spectra for 16 star forming galaxies at $z=2$ to 3. Referring to archival HST images, they obtained rotation curves for six galaxies. Although the angular resolution $\sim 0''.5$, corresponding to $\sim 5$ kpc, was not sufficient to resolve the details, the rotation velocities of 100 to 200 \kms have been observed, and the mean dynamical mass of the galaxies were shown to be greater than $4\times 10^{10}\Msun$.

 Genzel et al.(2008, 2011) obtained 2D H$\alpha$ velocity fields for star forming galaxies at $z=2-3$, obtaining rotation curves along the major axis for several objects. There is an increasing number observations of galaxies at $z\sim 1-3$ in near-infrared spectroscopy using the Hubble Space Telescope and large aperture telescopes (Epinat et  al. 2012;   Law et al. 2009;  Robertson and Bullock 2008; van der Wel and van der Marel 2008;  Shapiro et al. 2008). 
 
 The rotation curve for BzK 6004 at $z=2.387$ (Genzel et al. 2011) shows already similar rotation property to those in the nearby galaxies. Distant galaxies so far observed at redshifts $z<3$ appear to show no particular difference in rotation curve shapes and velocities from those of nearby galaxies. This may imply that the dynamical structure of spiral galaxies at the present time were already reached at these redshifts. Higher resolution rotation curves at higher redshifts would be crucial to investigate dynamical evolution of disk galaxies and their merging processes.
 
 { However, the inner rotation curves of high-redshift galaxies are still not precise enough to be compared with those of nearby galaxies. This is mainly due to the beam-smearing effect caused by the limited angular resolution of the instruments. This effect usually results in milder rise and slower velocity of rotation in the inner and nuclear regions of the galaxies, as well as in higher velocity dispersion, than the true values. It may be remembered that the angular resolution of $0.01''$ achieved by the current optical/infrared/sub-mm instruments corresponds to linear resolution of only $\sim 200$ pc at $z>2$ or at distances greater than several billion pc.
 }

  		\chapter{MASS DISTRIBUTION IN DISK GALAXIES}   

There are two major methods to measure the mass distribution using rotation curves, which are the direct method and the decomposition method. 
In the direct method, the rotation velocity is used to directly calculate the mass distribution.  
In the decomposition method, a galaxy is represented by superposition of several mass components, and the rotation curve is fitted by searching for the best-fitting parameters of the components.

\section{Direct Mass Determination}

By this direct method the mass distribution in a galaxy is calculated directly from the rotation curve. No functional form is necessary to be given a priori. Only an assumption has to be made, either if the galaxy's shape is a sphere or a flat disk. The 'true' mass profile is considered to lie between the two extreme cases of the spherical and axisymmetric disk distribution.  

\subsection{Spherical case}

On the assumption of spherical distribution, the mass inside radius $R$ is given by
\begin{equation}
M(R)=\frac{R {V(R)}^{2}}{G}.
\label{masssphere}
\end{equation}  
Then the surface-mass density (SMD) ${\Sigma}_{S}(R)$ at $R$ is calculated  by 
\be
\Sigma_{\rm S}(R) = 2 \int\limits_0^{\infty} \rho (r) dz .
\label{smdsphere}
\ee 
Remembering 
\begin{equation}
\rho(r) =\frac{1}{4 \pi r^2} \frac{dM(r)}{dr},
\label{rhosphere}
\end{equation}
the above expression can be rewritten as
\be
\Sigma_{\rm S}(R) = \frac{1}{2 \pi} \int\limits_R^{\infty} \frac{1}{r \sqrt{r^2-R^2}} \frac{dM(r)}{dr}dr .
\label{smdsphere_b}
\ee 
This gives good approximation for spheroidal component in the central region, but results in underestimated mass in the outer regions. Particularly, the approximation fails in the outermost region near the end of rotation curve due to the edging effect of integration.  

\subsection{Flat-disk case} 
  
The SMD of a flat thin disk, ${\Sigma}_{\rm D}(R)$, is calculated by solving the Poisson's equation on the assumption that the mass is as symmetrically distributed in a disk of negligible thickness (Freeman 1970; Binney  and  Tremaine 1987). It is given by
$$
{\Sigma}_{\rm D}(R) =\frac{1}{{\pi}^2 G} \times
$$
\be
\left[ \frac{1}{R} \int\limits_0^R 
{\left(\frac{dV^2}{dr} \right)}_x K \left(\frac{x}{R}\right)dx 
+ \int\limits_R^{\infty} {\left(\frac{dV^2}{dr} \right)}_x K \left
(\frac{R}{x}\right) \frac{dx}{x} \right].
\label{smdflat}
\end{equation}
Here, $K$ is the complete elliptic integral, which becomes very large when $x\simeq R$. 

A central black hole may influence the region within $\sim 1$, but it does not affect much the galactic scale SMD at lower resolution. Since it happens that there exist only a few data points in the innermost region, the reliability is lower than the outer region. Since the rotation curves are nearly flat or declining outward beyond maximum, the SMD values are usually slightly overestimated in the outer disk.

\subsection{Milky Way's direct mass calculation} 

\begin{figure} 
\begin{center}
\includegraphics[width=7cm]{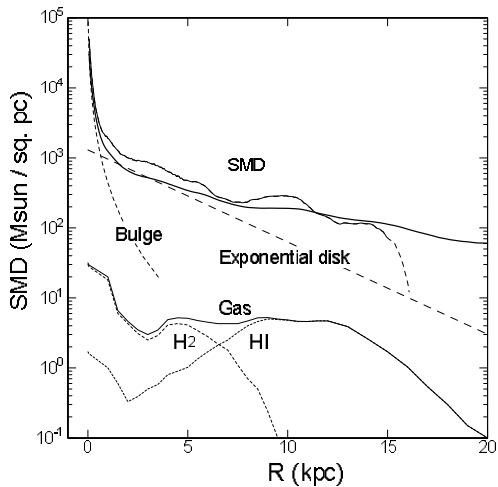}   
\end{center}
\caption{SMD distribution in the Milky Way. Thick line shows the directly calculated SMD for thin disk case, and thin line shows result for spherical case. The long dashed smooth lines are model profile for \dv\ and exponential disk. The lower lines show SMD of interstellar gas made by annulus-averaging of figure \ref{HIH2face} (Nakanishi and Sofue 2003, 2006, 2016). HI and H$_2$ gas SMDs are also shown separately by dotted lines. (Bottom) Same, but for spiral galaxies.}
\label{fig-smd-mw}      
\end{figure}

Figure \ref{fig-smd-mw}  shows SMD distributions in the Galaxy calculated by the direct methods for the sphere and flat-disk cases, compared with SMD calculated for the components obtained by deconvolution of the rotation curve. There is remarkable similarity between the results by direct methods, and by RC deconvolution. 

The SMD is strongly concentrated toward the center, reaching SMD as high as $\sim 10^5 \Msun~{\rm pc}^2$ within $\sim 10$ pc.  The galactic disk appears as an exponential disk as indicated by the straight-line at $R \sim 3 $ to  8 kpc on the semi-logarithmic plot. It is worth to note that the dynamical SMD is dominated by dark matter, because the SMD is the projection of extended dark halo. Note, however, that the volume density in the solar vicinity is dominated by disk's stellar mass.
The outer disk is followed by an outskirt with a slowly declining SMD profile, which indicates the dark halo and extends to the end of the rotation curve measurement.

In figure \ref{fig-smd-mw}, we also show the radial distribution of interstellar gas density, as calculated by azimuthally averaging the gas density distribution in figure \ref{HIH2face}. The gas density is much smaller than the dynamical mass density by an order of magnitude. The SMD of ISM is  $\sim 5.0 \msqpc$  $R\sim 8$ kpc, sharing only several percents of disk mass density $\sim 87.5 \msqpc$. 

\begin{figure} 
\begin{center}
\includegraphics[width=7cm]{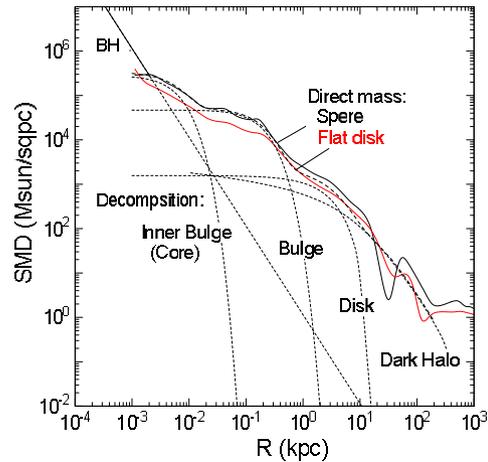}  
\end{center}
\caption{Directly calculated SMD of the Milky Way by spherical (black thick line) and flat-disk assumptions by log-log plot, compared with the result by deconvolution method (dashed lines). The straight line represents the black hole with mass $3.6\times 10^6 \Msun$.}
\label{smd-log}   
\end{figure}

The SMDs obtained by deconvolution and direct methods are consistent with each other within a factor of $\sim 1.5$. At $R\sim 8$ kpc, the directly calculated SMD is $\sim 300 \Msun~{\rm pc}^{-2}$, whereas it is $\sim 200-250 \Msun~{\rm pc}^{-2}$ by deconvolution, as shown in figure \ref{smd-log}.

\subsection{Spiral galaxies' direct mass}

\begin{figure*} 
\begin{center}      
\includegraphics[width=12cm]{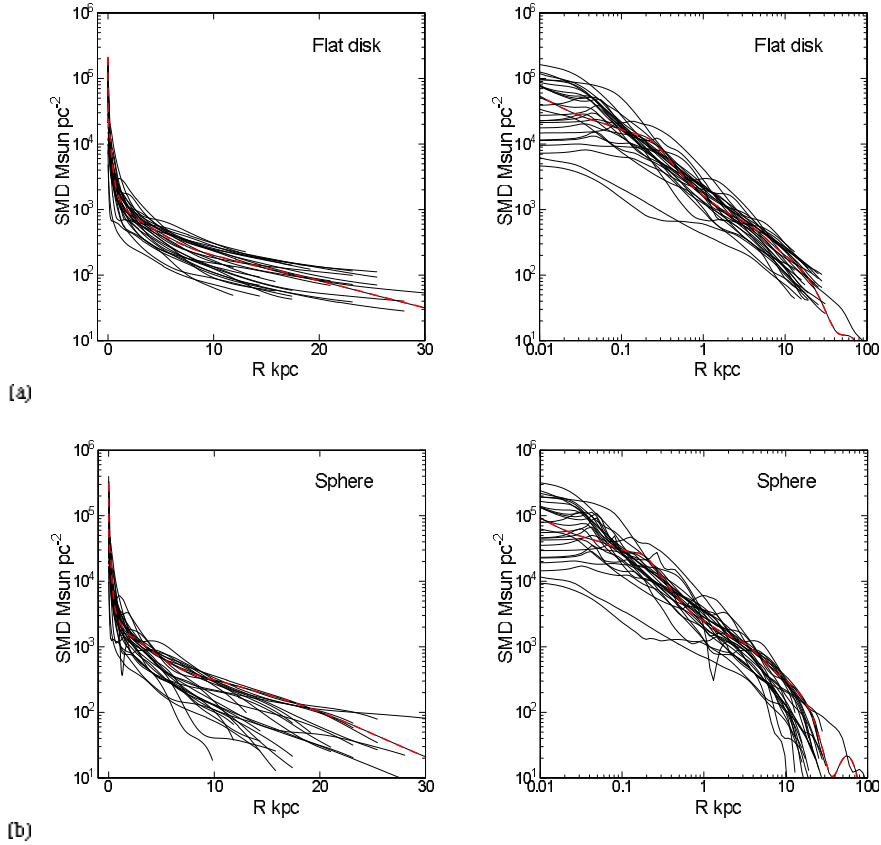} 
\end{center}
\caption{(a) Direct SMD in spiral galaxies with end radii of RC greater than 15 kpc (from Sofue 2016) calculated under flat disk assumption, and the same in 
logarithmic radius (Sofue 2016b). Red dashed lines indicates the Milky Way. (b) Same but under sphere assumption.}  
\label{fig-smd}   
\end{figure*}

Figure \ref{fig-smd} shows SMD distributions of spiral galaxies calculated for the rotation curves shown in figure \ref{allrc} using the direct methods. 
Results for flat-disk assumption give stable profiles in the entire galaxy, while the sphere assumption yields often unstable mass profile due to the edge effect in the outermost regions. On the other hand, the central regions are better represented by the sphere assumption because of the suspected spherical distribution of mass inside the bulge.

The calculated SMD profiles for galaxies are similar to that of the Milky Way.  Namely, dynamical structures represented by the density profiles in spiral galaxies are similar to each other, exhibiting universal characteristics as shown in the figures: high central concentration, exponential disk, and outskirt due to the dark halo. 

\subsection{Mass-to-luminosity ratio}

The farthest rotation velocity so far measured for a spiral galaxy is that for the Milky Way and M31 up to $\sim 300$ kpc, where the kinematics of satellite galaxies was used to estimate the circular velocities (Sofue 2012, 2013b). The obtained rotation curve was shown to be fitted by the NFW density profile. 

 { Note, however, that the NFW model predicts declining rotation only beyond  galacto-centric distances farther than $\sim 50$ kpc. Inside this radius, there is not much difference in the RC shapes of NFW model and isothermal model, predicting almost flat (NFW) or perfectly flat (isothermal) rotation. Practically for most galaxies with rotation curves up to $\sim 30$ kpc, both the models yield about the same result about their halos. In either models, observed rotation velocities in spiral galaxies show that the mass in their halos is dominated by dark matter.}

\begin{figure} 
\begin{center}
\includegraphics[width=7cm]{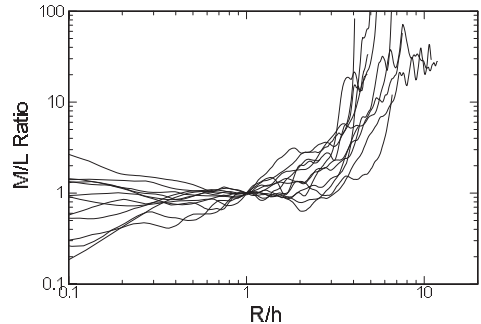}  \\ 
\end{center}
\caption{M/L ratios in spiral galaxies normalized at scale radii using data from Takamiya and Sofue (1999), where $L$ is the optical luminosity. 
   }
\label{fig-MLratio}  
\end{figure}

The mass-to-luminosity (M/L) ratio can be obtained by dividing SMD by surface luminosity profiles (Forbes 1992; Takamiya and Sofue 2000; Vogt et al. 2004a). Figure \ref{fig-MLratio} shows M/L ratios for various spiral galaxies normalized at their scale radii (Takamiya and Sofue 2000). While the M/L ratio is highly variable inside their disk radii, it increases monotonically toward galaxies' edges.{ However, the outermost halos farther than $\sim 20-30$ kpc, beyond the plotted radii in the figure, the M/L ratio is extremely difficult to determine from observations because of the limited finite radii of the luminous disks, beyond which the surface brightness is negligibly low, and hence the M/L ratio tends to increase to infinity.}

\section{Rotation Curve Decomposition into Mass Components}  

\subsection{Superposition of mass components}

The rotation velocity is written by the gravitational potential as
\be
V(R)=\sqrt{R  {\partial \Phi \over \partial R}},
\ee
where
\be
\Phi=\Sigma  \Phi_i
\ee
with $\Phi_i$ being the potential of the $i$-th mass component. Knowing that $V_i(R)=R\partial \Phi_i / \partial R$, we have
\be
V(R)=\sqrt{\Sigma V_i^2}. 
\ee
It is often assumed that there are multiple components that are the central black hole, a bulge, disk and a dark halo:
\be
V(R)= \sqrt{V_{\rm BH}(R)^2+V_{\rm b}(R)^2+V_{\rm d}(R)^2+V_{\rm h}(R)^2}.
\ee 
Here, the suffice BH represent black hole, b for bulge, d for disk and h for the dark halo. 

{ The parameters (masses and scale radii of individual components) are iteratively determined by the least $\chi^2$ fitting (Carignan 1985; Carignan and Freeman 1985). First, an approximate set of the parameters are given as an initial condition, where the values of the parameters are hinted by luminosity profiles of the bulge and disk, and by the shape and amplitude of the rotation curve for the dark halo. All the parameters are fitted at once to fit the observed rotation curve.

When precise rotation curves are available with a larger number of data points, particularly in the resolved innermost regions of the Milky Way and nearby galaxies, the fitting may be divided into several steps according to the components in order to save the computation time (Sofue 2012). The time per one iteration is proportional to $nN$, where $n$ is the number of data points and $N$ is the number of combination of parameters. The combination number is given by $N=(\Sigma_i n_i)!$, where $n_i$ is the number of parameters of the $i$-th component. On the other hand, it is largely reduced to $N=\Sigma_i (n_i !)$ in the step-by-step method. 

Also, it is useful to divide the fitting radii depending on the component's properties. Namely, a black hole and bulge may not be fitted to data beyond the disk, e.g., beyond $R\sim 10$ kpc, while a halo may not be fitted at $<\sim 1$ kpc. This procedure can save not only the time, but also the degeneracy problem (Bershady et al. 2010a,b). Degeneracy happens for data with low resolution and small number of measurements, often in old data, in such a way that a rotation curve is represented equally by any of the mass models, or, the data is fitted either by a single huge bulge or disk, or by a single tiny halo.

In the step-by-step method, a set of the parameters are assumed, first, as the initial condition as above. The fitting is started individually from the innermost component having the steepest rise (gradient), which is particularly necessary when a black hole is included. Next, the steeply rising part by the bulge is fitted, and then gradual rise and flat parts are fitted by the disk. Finally, the residual outskirt is fitted by a dark halo. This procedure is repeated iteratively, starting again from the innermost part, until the $\chi^2$ value is minimized. }

Observed rotation curves in spiral galaxies may be usually fitted by three components of the bulge, disk and dark halo. In the Milky Way, the rotation curve may better be represented by four or five components which are the central black hole, multiple bulges, an exponential disk, and dark halo.

\begin{figure} 
\begin{center}
\includegraphics[width=5.5cm]{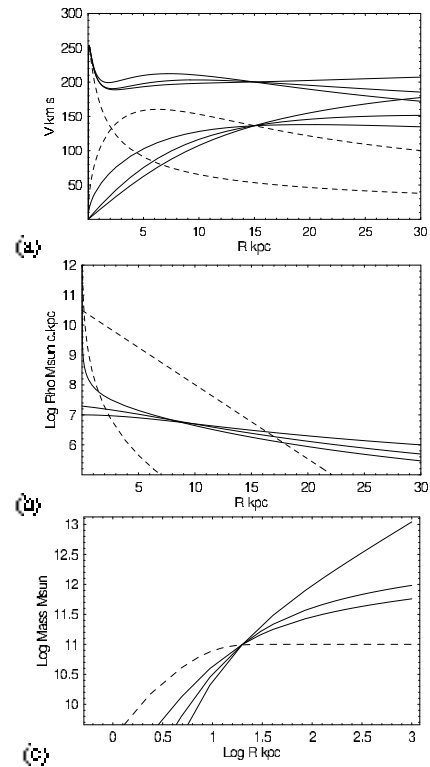}  
\end{center}
\caption{(a) Analytic rotation curve composed of bulge, disk and dark halo components represented by isothermal, Burkert (1995) and NFW models (full lines from top to bottom at $R=30$ kpc). Dashed lines represent \dv\ bulge and exponential disk.
(b) Corresponding volume densities.
(c) Corresponding enclosed mass within radius r.}
\label{fig-dh}   
\end{figure}

\subsection{Massive black hole}   

The Galactic Center of the Milky Way is known to nest a massive black hole of mass of $M_{\rm BH}= 2.6-4.4 \times 10^6\Msun$ (Genzel et al. 1994, 1997, 2000, 2010),  $4.1-4.3\times 10^6\Msun$ (Ghez et al. 1998, 2005, 2008), and $3.95 \times 10^6\Msun$ (Gillessen et al. 2009).  
  
A more massive black hole has been observed in the nucleus of the spiral galaxy NGC 4258 with the mass of $3.9 \times 10^{7}\Msun$ (Nakai et al. 1993;   Miyoshi et al. 1995; Herrnstein et al. 1999) from VLBI observations of the water maser lines. Some active galaxies have been revealed of rapidly rotating nuclear torus of sub parsec scales indicating massive black holes in several nearby active galactic nuclei, and  increasing number of evidences for massive black holes in galactic nuclei have been reported (Melia 2010).   
Statistics shows that the mass (luminosity) of bulge is positively related to the mass of central black hole (Kormendy  and  Westpfahl 1989; Kormendy 2001).     
\subsection{\dv\ bulge}   

\begin{figure}
\begin{center}
\includegraphics[width=7cm]{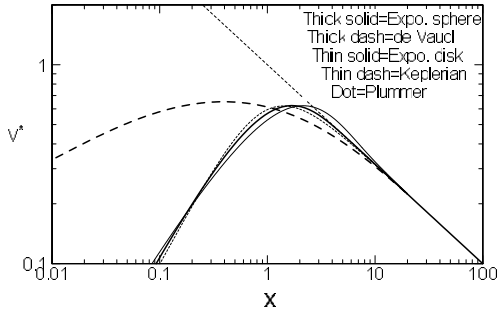}   
\end{center}
\caption{Comparison of normalized rotation curves for the exponential spheroid, de Vaucouleurs spheroid, and other typical models, for a fixed total mass. The exponential spheroid model is almost identical to that for Plummer law model. The central rise of de Vaucouleurs curve is proportional to $\propto r^{1/2}$, while the other models show central velocity rise as $\propto r$.}
\label{Dfunc}
\end{figure}

The most commonly used profile to represent the central bulge is the \dv\ (1958) law (figure \ref{Dfunc}), which was originally expressed by a surface-brightness distribution at projected radius $R$ by
\be
{\rm log} \beta = - \gamma(\alpha^{1/4}-1),
\label{eq-dv}
\ee
where $\gamma=3.3308$. Here, $ \beta=B_{\rm b}(R)/B_{\rm be}$,  $ \alpha=R/R_{\rm b} $, and $B_{\rm b}(R)$ is the surface-brightness normalized by the value at radius $R_{\rm b}$, $B_{\rm be}$.  

The same \dv\ profile for the surface mass density is usually adopted for the surface mass density as
\be \Sigma_{\rm b}(R)=\lambda_{\rm b} B_{\rm b}(R)= \Sigma_{\rm be} {\rm exp} \left[-\kappa \left(\left(R \over R_{\rm b} \right)^{1/4}-1\right)\right]
\label{eq-smdb}
\ee 
with $ \Sigma_{\rm bc} = 2142.0 \Sigma_{\rm be} $ for $\kappa=\gamma {\rm ln} 10=7.6695$.  
Here, $\lambda_{\rm b}$ is the M/L ratio assumed to be constant.  

Equations (\ref{eq-dv}) and (\ref{eq-smdb}) show that the central SMD at $R=0$ attains a finite value, and the SMD decreases steeply outward near the center. However, the decreasing rate gets much milder at large radii, and the SMD decreases slowly, forming an extended outskirt (\dv 1958). 

The cylindrical mass inside $R$ is calculated
\be M_{\rm b:cyl}(R)= 2 \pi \int_0^R x \Sigma_{\rm b}(x) dx.
\label{Mbcyl}
\ee
Total mass of the bulge given by
\be M_{\rm bt}= 2 \pi \int_0^\infty R \Sigma_{\rm b}(R) dR =\eta R_{\rm b}^2 \Sigma_{\rm be}
\ee
with $\eta=22.665$. A half of the total projected (cylindrical) mass is equal to that inside a cylinder of radius $R_{\rm b}$.

The volume mass density $\rho(r)$ at radius $r$ for a spherical bulge is now given by 
\be
\rho(r) = {1 \over \pi} \int_r^{\infty} {d \Sigma_b(x) \over dx} {1 \over \sqrt{x^2-r^2}}dx.
\label{eq-rhob}
\ee 
The circular velocity is thus given by the Kepler velocity of the mass inside $R$ as
\be
 V_{\rm b}(R) = \sqrt{ GM_{\rm b:sph}(R) \over R} ,
 \ee
 where
 \be
 M_{\rm b:sph}(R)=4\pi \int_0^R r^2\rho(r)dr.
\label{Mbsph}
 \ee
 Note that the spherical mass $ M_{\rm b:sph}(R)$ is smaller than the cylindrical mass $M_{\rm b:cyl}(R)$ given by equation (\ref{Mbcyl}). 
At large radii, the velocity approximately decreases by Keplerian-law.  Figure \ref{Dfunc}) shows the variation of circular velocity for a \dv\ bulge.

The \dv\ law has been extensively applied to fit spheroidal components of late type galaxies (Noordermeer 2007, 2008). S\'ersic (1958) has modified the law to a more general form  $e^{-(R/r_e)^n}$. The \dv\ and S\'ersic laws were fully discussed in relation to its dynamical relation to the galactic structure based on the more general profile (Ciotti 1991; Trujillo 2002). 
The \dv\ law has been also applied to fit the central rotation curve of the Milky Way as shown in figure \ref{logrc_mw}. However, it was found that the \dv law cannot reproduce the observations inside $\sim 200$ pc. 

It is interesting to see the detailed behavior of the \dv\ law. At the center, $\Sigma$ reaches a constant, and volume density varies as $\propto 1/r$, leading to circular velocity $V \propto r^{1/2}$ near the center. Thus, the rotation velocity rises very steeply with infinite gradient at the center. It should be compared with the mildly rising velocity as $V \propto r$ in the other models. Figure \ref{Dfunc} compares normalized behaviors of rotation velocity for \dv and other models. 

The \dv rotation curve shows a much broader maximum in logarithmic plot compared to the other models, because the circular velocity rises as $V \propto \sqrt{M(<r)/r} \sim \sqrt r$. The particular behavior can be better recognized by comparing the half-maximum logarithmic velocity width with other models, where the width is defined by
\be
\Delta_{\rm log}={\rm log} ~{r_2 \over r_1}.
\ee
Here, $r_2$ and $r_1$ ($r_2>r_1$) are the radii at which the rotational velocity becomes half of the maximum velocity. From the figure, we obtain $\Delta_{\rm log} = 3.0$ for \dv, while $\Delta_{\rm log}=1.5$ for the other models as described later. Thus the \dv's logarithmic curve width is twice the others, and the curve's shape is much milder.

\subsection{Other bulge models}

Since it was shown that the \dv law fails to fit the observed central rotation, another model has been proposed, called the exponential sphere model. In this model, the volume mass density $\rho$ is represented by an exponential function of radius $r$ with a scale radius $a$ as
\be
\rho(r)=\rho_{\rm c} e^{-r/a}.
\ee
The mass involved within radius $r$ is given by
\be
M(R)=M_0 F(x),
\ee
where $x=r/a$ and
\be
F(x)=1-e^{-x}(1+x+x^2/2).
\ee
The total mass is given by
\be
M_0=\int_0^\infty 4 \pi r^2 \rho dr=8 \pi a^3 \rho_{\rm c}.
\ee
The circular rotation velocity is then calculated by
\be
V(r)=\sqrt{G M/r}=\sqrt{{G M_0\over a} F\left({r\over a}\right)}.
\ee 
In this model the rotational velocity has narrower peak near the characteristic radius in logarithmic plot as shown in figure \ref{Dfunc}. Note that the exponential-sphere model is nearly identical to that for the Plummer's law, and the rotation curves have almost identical profiles. In this context, the Plummer law can be used to fit the central bulge components in place of the present models.

\subsection{Exponential disk }

The galactic disk is represented by an exponential disk (Freeman 1970), where the surface mass density is expressed as
\be 
\Sigma_{\rm d} (R)=\Sigma_{dc} {\rm exp}(-R/R_{\rm d}).
\label{eq-smdd}
\ee 
Here, $\Sigma_{dc}$ is the central value, $R_d$ is the scale radius. The total mass of the exponential disk is given by $M_{\rm disk}= 2 \pi \Sigma_{dc} R_{\rm d}^2$.
The rotation curve for a thin exponential disk is expressed by (Binney and Tremaine 1987).  
\be
V_{\rm d(R)}=\sqrt{R{\partial \Phi \over \partial R}}
\label{expdisk}
\ee
$$
=\sqrt{4 \pi G \Sigma_0 R_{\rm d}y^2[I_0(y)K_0(y)-I_1(y)K_1(y)]},
$$
where $y=R/ (2R_{\rm d} $, and $I_i$ and $K_i$ are the modified Bessel functions.

If the surface mass density does not obey the exponential law, the gravitational force $f(R)$ can be calculated by integrating the $x$ directional force caused by mass element $\Sigma_{\rm d}' (x) dx dy$ in the Cartesian coordinates $(x,y)$:
\be 
f(R)=G \int_{-\infty}^\infty \int_{-\infty}^\infty 
{\Sigma_d (x) (R-x) \over s^3} dx dy ,
\ee
where $ s=\sqrt{(R-x)^2+y^2} $. The rotation velocity is then given by 
\be
V_d(R) = \sqrt{f R}.
\ee
This formula can be used for any thin disk with an arbitrary SMD distribution $\Sigma(x,y)$, even if it includes non-axisymmetric structures.

\subsection{Isothermal and NFW dark halos}

\subsubsection{Evidence for dark matter halo}
The existence of dark halos in spiral galaxies has been firmly evidenced from the well established difference between the galaxy mass predicted by the luminosity and the mass predicted by the rotation velocities (Rubin et al. 1980-1985; Bosma 1981a, b; Kent 1986, 1987;  Persic  and  Salucci 1990; Salucci  and  Frenk 1989; Forbes 1992;  Persic et al. 1996; H\'{e}raudeau  and  Simien 1997; Takamiya  and  Sofue 2000).  

In the Milky Way, extensive analyses of motions of non-disk objects such as globular clusters and dwarf galaxies in the Local Group have shown flat rotation up to $\sim 100$ kpc, and then declining rotation up to $\sim 300$ kpc (Sofue 2013, 2015). The outer rotation velocities have made it to analyze the extended distribution of massive dark halo, which is found to fill the significantly wide space in the Local Group. Bhattacharjee et al.(2013, 2014) have analyzed non-disk tracer objects to derive the outer rotation curve up to 200 kpc in order to constrain the dark matter mass of the Galaxy, reaching a consistent result. As will be shown later, the dark halos of the Milky Way and M31 are shown to be better represented by the NFW model than by isothermal halo model.

\subsubsection{Isothermal halo} 

The simplest model for the flat rotation curve is the semi-isothermal spherical distribution (Kent 1986; Begeman et al. 1991), where the density is written as
\be
\rho_{\rm iso} (R)={\rho_{\rm iso} ^0 \over  1+ (R/h)^2}, 
\label{eq_iso} 
\ee
where $\rho_{\rm iso}$ and $h=R_{\rm h}$ are the central mass density and scale radius, respectively. The circular velocity is given by 
\be 
V_h(R)=V_\infty \sqrt{1-\left(h \over R \right) {\rm tan}^{-1}\left(R \over h \right) },  
\ee
which approaches a constant rotation velocity $V_\infty$ at large distances. 
At small radius, $R\ll h$, the density becomes nearly constant equal to $\rho_{\rm iso}^0$ and the enclosed mass increases steeply as $M(R) \propto R^3$. At large radii, the density decreases with $\rho_{\rm iso} \propto R^{-2}$. The enclosed mass increases almost linearly with radius as $M(R) \propto R$.

\subsubsection{Navarro-Frenk-White (NFW) model}

The most popular model for the dark halo is the NFW model (Navarro, Frenk and White 1996, 1997) empirically obtained from numerical simulations in the cold-dark matter scenario of galaxy formation. Burkert's (1995) modified model is also used when the singularity at the nucleus is to be avoided. The NFW density profile is written as 
\be 
\rho_{\rm NFW} (R)={\rho_{\rm NFW} ^0 \over (R/h)/[1+(R/h)^2]}.
\label{eq-nfw} 
\ee 
The circular velocity is equal to
\be
V_h (R)=\sqrt{GM_{\rm h} (R)\over R},
\ee
where $M_h$ is the enclosed mass within the scale radius $h$.

At $R \ll h$, the NFW density profile behaves as $\rho_{\rm NFW}\propto 1/R$, yielding an infinitely increasing density toward the center, and the enclosed mass behaves as $M(R) \propto R^2$. The Burkert's modified profile tends to constant density $\rho_{\rm Bur}^0$, similar to the isothermal profile.

At large radius the NFW shows density profile as $\rho_{\rm NFW,~Bur} \propto R^{-3}$, yielding logarithmic increasing of mass, $M(R) \propto {\rm ln}~ R$ (Fig. \ref{fig-dh} ). Figure \ref{fig-dh} shows density distributions for the isothermal, NFW and Burkert models.  

\subsection{Plummer and  Miyamoto-Nagai potential}

The current mass models described above do not necessarily satisfy the Poisson's equation, and hence, they are not self consistent for representing a dynamically relaxed self-gravitating system. In order to avoid this inconvenience, Plummer-type potentials are often employed to represent the mass distribution.  

The Miyamoto and Nagai's (MN) (1975) potential is one of the most convenient Plummer-type formulae to describe a disk galaxy's potential and mass distribution in an analytic form:
\begin{equation}
 \Phi  
 = \sum_{i=1}^n {-GM_i \over \sqrt{R^2+(a_i+\sqrt{z^2+b_i^2})^2}},
 \label{eqMN}
\end{equation}
where, $M_i$, $a_i$ and $b_i$ are the mass, scale radius and height of the $i$-th spheroidal component. The rotation velocity in the galactic plane at $z=0$ is given by
\be
 V_{\rm rot}(R)   
 =R\sqrt{\sum_{i=1}^n {GM_i \over [R^2+(a_i+b_i)^2]^{3/2}}}.
\label{eqVrot} 
\ee
The mass distribution is given by the Poisson's equation:
$$ 
\rho(R, z)=-{\Delta \Phi \over 4 \pi G}
={1 \over 4 \pi}\sum_{i=1}^n b_i^2 M_i  
$$
\be
\times {a_iR^2+(a_i+3\sqrt{z^2+b_i^2})(a_i+\sqrt{z^2+b_i^2})^2
\over
[R^2+(a_i+\sqrt{z^2+b_i^2})^2]^{5/2}(z^2+b_i^2)^{3/2}}. 
\label{eqMNden} 
\ee

This model was used to approximate the observed rotation curve of the Milky Way (Miyamoto and Nagai 1975). Their proposed parameters are often used to represent a bulge and disk for numerical simulations not only in the original form, but also by adding a larger number of components by choosing properly the masses and scale radii. The original parameters were given to be 
$M_{\rm bulge} \sim  2.05\times 10^{10}\Msun,~ a_{\rm bulge}=0, ~ b_{\rm bulge}\sim 0.495$ kpc, 
and 
$M_{\rm disk} \sim 2.547\times 10^{11} \Msun,~ a_{\rm disk}\sim 7.258~{\rm pc}, ~b_{\rm disk}\sim 0.520$ kpc. 

\subsection{Degeneracy problem}

The models described here are expressed by single-valued simple analytic functions. Each of the functions has only two parameters (mass and size). The black hole is expressed by the Keplerian law with only one parameter (mass). The mass and scale radius of the \dv\ law are uniquely determined, if there are two measurements of velocities at two different radii, and so for the exponential disk and NFW halo. 

If the errors are sufficiently small, the three major components (bulge, disk and halo) can be uniquely determined by observing six velocity values at six different radii. In actual observations, the number of data points are much larger, while the data include measurement errors. Hence, the parameters are determined by statistically by applying the least-squares fitting method or the least $\chi^2$ method. Thereby, the most likely sets of the parameters are chosen as the decomposition result.

However, when the errors are large and the number of observed points are small, degeneracy problem becomes serious (Bershady et al. 2010a,b), where the fitting is not unique. Such a peculiar case sometimes happens, when the rotation curve is mildly rising and tends to flat end, that the curves fitted in almost the same statistical significance either by a disk and halo, by a single disk, or by a single halo. Such mild rise of central rotation is usually observed in low-resolution measurements.

\section{Rotation Curve Decomposition in the Milky Way}

Observed mass components in the Galaxy and spiral galaxies are often expressed by empirical functions derived by surface photometry of the well established bulge and disk. Also, central black hole and outermost dark halo are the unavoidable components to describe resolved galactic structures. 

\subsection{Black hole, bulge, disk and halo of the Milky Way}

In the rotation curve decomposition in the Milky Way, the following components have been assumed (Sofue 2013):\\
(1) The central black hole with mass $M_{\rm BH}=4\times 10^6 \Msun$,\\
(2) An innermost spheroidal component with exponential-sphere density profile, or a central massive core,\\
(3) A spheroidal bulge with an exponential-sphere density profile, and\\
(4) An exponential flat disk.\\
(5) A dark halo with NFW profile. \\
An initially given approximate parameters were adjusted to lead to the best-fitting values using the least $\chi^2$ method. Figure \ref{logrc_mw} shows the fitted rotation curve, which satisfactorily represents the entire rotation curve from the central black hole to the outer dark halo. 

The fitting of the two peaks of rotation curve at $r\sim 0.01$ kpc and $\sim 0.5$ kpc are well reproduced by the two exponential spheroids. The figure also demonstrates that the exponential bulge model is better than the \dv model. 
It must be mentioned that the well known \dv profile cannot fit the Milky Way's bulge, while it is still a good function for fitting the bulges in extragalactic systems. Table \ref{tab_milkyway} lists the fitting parameters for the Milky Way.  

\begin{figure} 
\begin{center}    
\includegraphics[width=7cm]{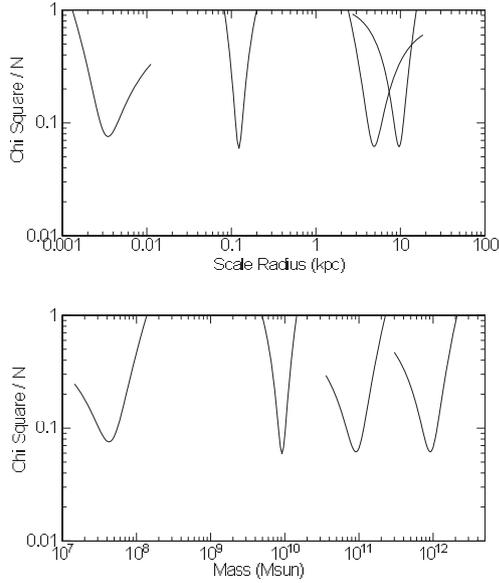}      
\end{center}
\caption{$\chi^2$ plots around the best-fit values of the scale radii of the deconvolution components to obtain figure \ref{logrc_mw},
} 
\label{mw_chi}  
\end{figure}

\begin{figure}
\begin{center}    
\includegraphics[width=3.5cm]{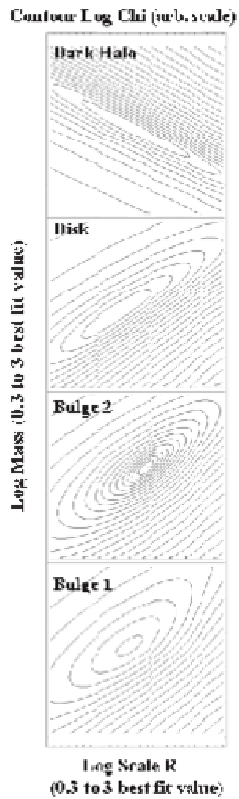}  
\end{center}
\caption{Contour presentation of two dimensional distribution of log $\chi^2$ value around the best-fit points in the scale radius-mass space to obtain figure \ref{logrc_mw}. } 
\label{mw_chimap} 
\end{figure}

\begin{table} 
\caption{Parameters for the mass components of the Galaxy (Sofue 2013b) $^\dagger$} 
\begin{tabular}{ll}  
\hline\hline  
Mass component & Quantities  \\   
\hline   
Black hole 
       & $M_{\rm bh}=3.6\times 10^6\Msun^\ddagger$  \\ 
\hline
Bulge 1 (massive core)& $\ab= 3.5 \pm 0.4  $ pc  \\ 
       & $\Mb= 0.4 \pm 0.1\times 10^8\Msun$  \\ 
\hline
Bulge 2 (main bulge)$^*$ & $\ab= 120  \pm 3  $ pc  \\ 
       & $\Mb= 0.92 \pm 0.02\times 10^{10}\Msun$ \\  
\hline
Disk   & $\ad= 4.9 \pm  0.4  $ kpc \\
       & $\Md=0.9 \pm 0.1 \times 10^{11}\Msun$ \\  
\hline
Dark halo 
       & $h= 10 \pm 0.5 $ kpc \\ 
       & $\rho_0= 2.9 \pm 0.3 \times 10^{-2} \Msun{\rm pc}^{-3}$ \\ 
       & $M_{R<200 {\rm kpc}}=0.7 \pm 0.1 \times 10^{12}\Msun$   \\ 
       & $M_{R<385{\rm kpc}}= 0.9 \pm 0.2 \times 10^{12}\Msun$   \\  
\hline
DM density at Sun 
       & $\rho_8 = 0.011 \pm 0.001 \Msun{\rm pc}^{-3}$  \\  
       & $ = 0.40 \pm 0.04 {\rm GeV~cm}^{-3}   $ \\  
\hline
\end{tabular} \\ 
 $^\dagger$ The adopted galactic constants are $(R_0,V_0)=$(8 kpc, 238 \kms) (Honma et al 2012).\\
    $^\ddagger$ Genzel et al. (2000 - 2010)\\
    $^*$ $\Mb$ is the surface mass enclosed in a cylinder of radius $\ab$, but not a spherical mass. 
\label{tab_milkyway}
\end{table}


\subsection{The Galactic Center}

\begin{figure}
\begin{center}
\includegraphics[width=7cm]{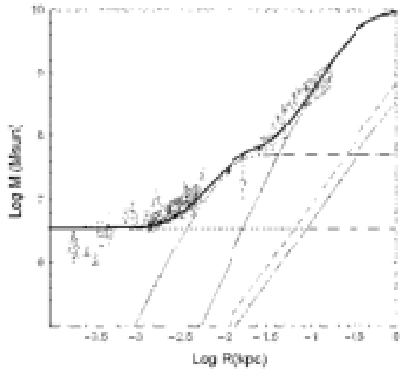}     
\end{center}
\caption{Enclosed mass calculated for the Galaxy's rotation curve compared to those by Genzel et al. (1994). The horizontal line, thin lines, and dashed line indicate the black hole, inner bulge, main bulge, disk, and dark matter cusp.}
\label{mass}
\end{figure}

Figure \ref{mass} shows the enclosed mass in the Galactic Center as a function of radius calculated for the parameters obtained by rotation curve decomposition. The plots are compared with the measured values at various radii, as compiled by Genzel et al. (1994), where their data have been normalized to $R_0=8.0$ kpc. It should be stressed that both the plots are in good agreement with each other.
The figure shows that the bulge density near the center tends to constant, so that the innermost enclosed mass behaves as $\propto r^3$ as the straight part of the plot indicates. On the other hand, the disk has a constant surface density near the center, and hence the enclosed mass varies as $\propto r^2$, as the straight line for the disk indicates. The NFW model predicts a high density cusp near the center with enclosed surface mass $\propto r^2$, as shown by the dashed line, exhibiting similar behavior to the disk.

\subsection{Local dynamical values and dark matter density}

The local value of the volume density of the disk can be calculated by $\rho_{\rm d}=\Sigma_{\rm d} /(2 z_0)$, where $z_0$ is the vertical scale height at radius $R=R_0$, when the disk scale profile is approximated by $\rho_{\rm d}(R_0, z)=\rho_{\rm d0}(R_0) {\rm sech} (z/z_0)$. The disk thickness has been observed to be $z_0=144\pm 10$ pc for late type stars using the HYPPARCOSS catalogue (Kong and Zhu 2008), while a larger value of 247 pc is often quoted as a traditional value (Kent et al. 1991). 

The local volume density by the bulge and dark halo is $\sim 10^{-4}$ and  $\sim 10^{-2}$ times the disk density, respectively. However, the SMD projected on the Galactic plane of the bulge contributes to 1.6\% of the disk value. It is interesting to note that the SMD of the dark halo exceeds the SMD of the disk by several times. 

The NFW model was found to fit the grand rotation curve quite well (Sofue 2012), including the declining part in the outermost rotation at $R\sim 40-400$ kpc. The local dark matter density is a key quantity in laboratory experiments for direct detection of the dark matter. 
Using the best fit parameters for the NFW model, the local dark matter density in the Solar neighborhood is calculated to be $\rho_0^\odot= 0.235\pm 0.030$ GeV cm$^{-3}$. This value may be compared with the values obtained by the other authors as listed in table \ref{tab_localdm}.

\begin{table} 
\caption{Local dark matter density $\rho_{\rm s}$ near the Sun in the Galaxy from the literature.} 
\begin{tabular}{ll}
\hline
Author &   $\rho_{\rm s}$ (GeV cm$^{-3}$) \\ 
\hline
Weber and de Boer (2010) &   0.2 - 0.4 \\   
Salucci et al. (2012) &   $ 0.43 \pm 0.10$  \\
Bovy and Tremaine (2012) &   $0.3\pm 0.1$ \\
Piffl et al. (2014) &   0.58 \\ 
Sofue (2013b), $V_0=200$ \kms $$ & $0.24\pm 0.03 $ \\
-----, $V_0=238$ \kms &   $0.40 \pm 0.04$ \\
Pato et al (2015a,b), $V_0=230$ \kms &$0.42\pm$ 0.25 \\
\hline
\end{tabular} 
\label{tab_localdm}   
\end{table}

\section{Decomposition of Galaxies' Rotation Curves}

\subsection{Bulge, disk and NFW halo decomposition of spiral galaxies}

The mass decomposition has been also applied extensively to spiral galaxies with relatively accurate rotation curves. It is a powerful tool to study the relations among the scale radii and masses of the bulge and disk with those of the dark halo, which provides important information about the structure formation in the universe (Reyes et al. 2012; Miller et al. 2014; Behroozi et al. 2013). 

Decomposition has been applied to galaxies as compiled in figure \ref{allrc} (Sofue 2016) by adopting the \dv, exponential and NFW density profiles for the bulge, disk and dark halo, respectively. The best-fitting values were obtained by applying the least $\chi^2$ method for $\Mb,~ \ab,~ \Md,~ \ad,~ \rho_0$ and $h$. The critical dark halo radius, $R_{200}$,  critical mass, $M_{200}$, as well as the mass, $M_h$, enclosed within the scale radius $h$ were also calculated, which are defined by
\be
M_{200}=200 \rho_{\rm c}{4\pi \over 3} R_{200}^3,
\label{mhalocrit}
\ee
and
\be
\rho_{\rm c}=3H^2_0/8\pi G,
\ee
with $H_0=72 $ \kms Mpc$^{-1}$ being the Hubble constant. 

\begin{figure}
\begin{center}
\includegraphics[width=7cm]{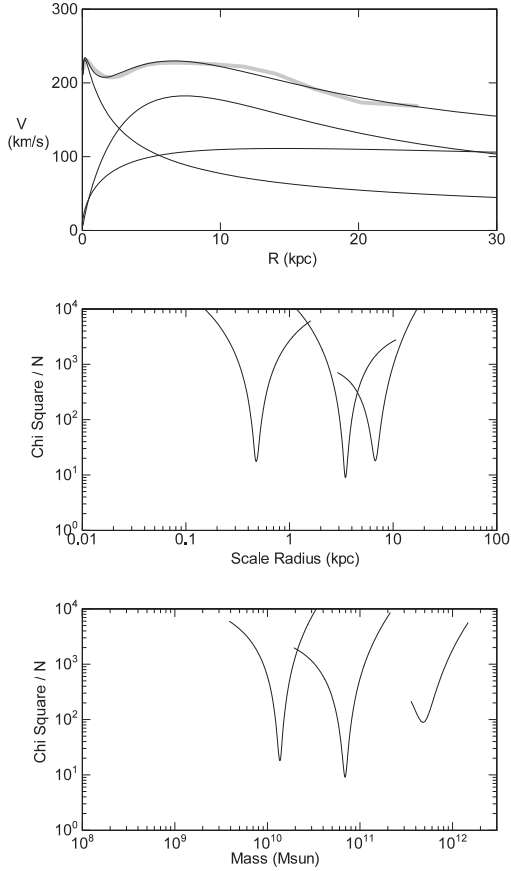}  
\end{center}
\caption{Rotation curve and $\chi^2$ fitting result for NGC 891 (top), distribution of $\chi^2/N$ around the best fit scale radii (middle) and masses (bottom) of the three components.  
} 
\label{rc} 
\end{figure}

Figure \ref{rc} shows examples of rotation curves and fitting result for NGC 891. The figure also shows the variation of $\chi^2$ values plotted against the parameters.  
Applying a selection criterion, the fitting was obtained for 43 galaxies among the compiled samples in figure \ref{allrc}, and the mean values of the results are shown in  table \ref{meanpara}.

\begin{table}  
\caption{Mean parameters for selected galaxies (Sofue 2016).} 
\begin{tabular}{lll}
\hline\hline  
Bulge size  &$\ab$& 1.5$\pm 0.2 $ kpc\\
--- mass & $\mb$& 2.3$\pm 0.4 \ \mten$\\ 
Disk size & $\ad$& 3.3$\pm 0.3$ kpc\\
--- mass  & $\md$& 5.7$\pm 1.1 \  \mten$\\ 
DH scale size  & $h$ & 21.6$\pm 3.9$ kpc\\
---mass within $h$ & $\mh$& 22.3$\pm 7.3\ \mten$ \\
---critical radius & $\rhalo$ & 193.7$\pm 10.8$ kpc\\
---critical mass & $\mhalo$ & 127.6$\pm 32.0\ \mten$\\ 
Bulge+Disk & $M_{\rm b+d} $ &  7.9$\pm 1.2\ \mten$\\ 
Bulge+Disk+Halo & $M_{\rm Total}\dagger$ & 135.6$\pm 32.0 \ \mten$\\
 \hline
B+D / Halo ratio &$M_{\rm b+d}/M_{\rm 200}$&$0.062\pm 0.018$\\ 
B+D / Total ratio & $M_{\rm b+d}/M_{\rm Total}$ & $0.059\pm 0.016$\\
 \hline
$\dagger\ M_{\rm Total}=M_{\rm 200+b+d}$ \\
\end{tabular}  
\label{meanpara} 
\end{table}

\subsection{Size and mass fundamental relations}

\begin{figure} 
\begin{center}
\includegraphics[width=60mm]{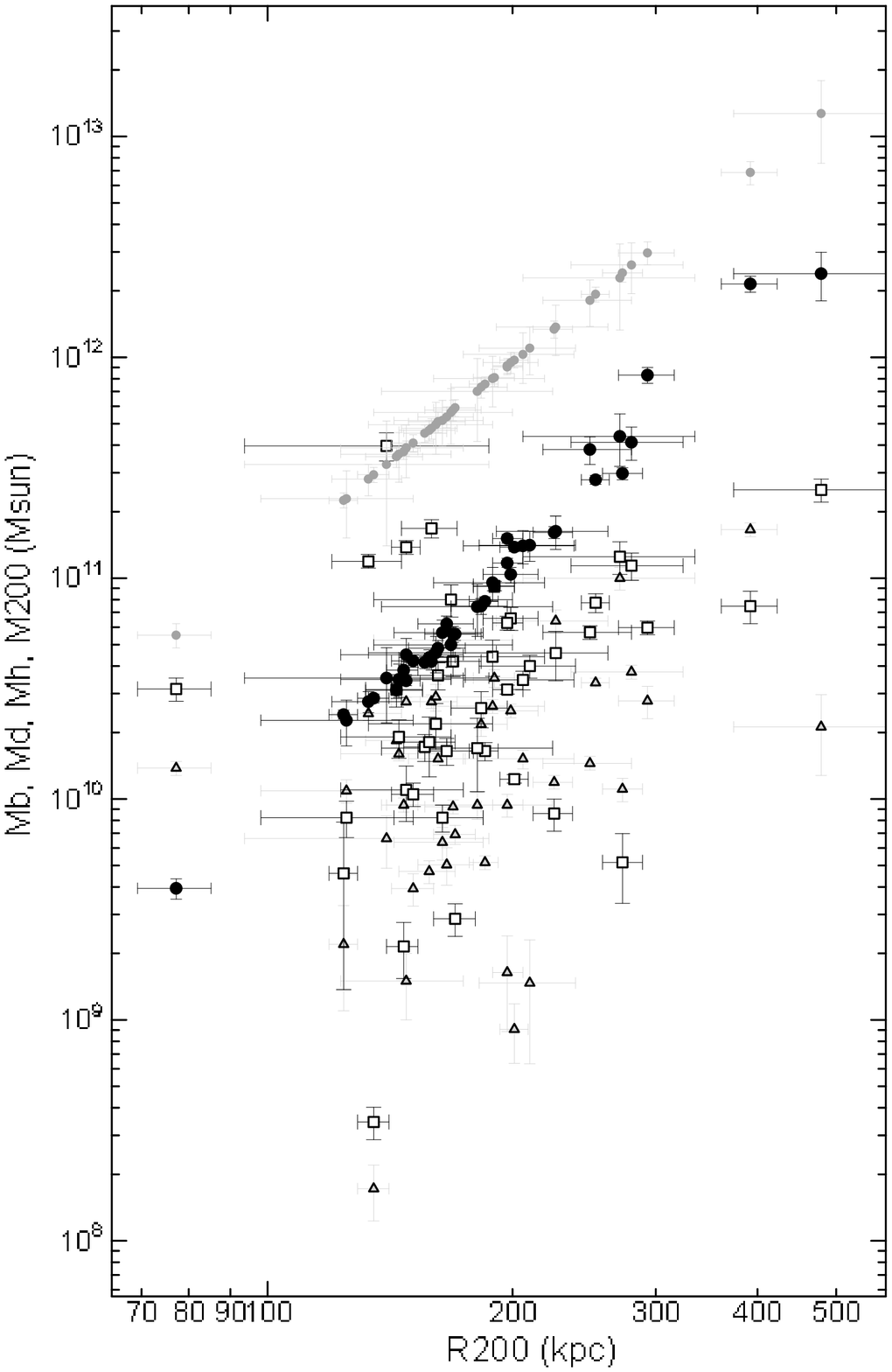} 
\end{center}
\caption{ Size-Mass relations for $(\rhalo,\mb)$ (triangles); $(\rhalo,\md)$ (rectangles);  $(\rhalo,\mh)$ (black dots);  $(\rhalo,\mhalo)$ (grey small dots) showing the trivial relation by definition of the critical mass. }  
\label{smrelation}  
\end{figure}

Correlations among the deconvolved parameters are useful to investigate the fundamental relations of dynamical properties of the mass components (e.g., Vogt et al. 2004a,b). Figure \ref{smrelation} shows plots of $\ab,~ \ad,$ and $h$ against the critical radius $\rhalo$ for the compiled nearby galaxies (Sofue 2015). It is shown that the bulge, disk and halo scale radii are positively correlated with $\rhalo$. Note that the tight correlation between $h$ and $R_{200}$ includes the trivial internal relation due to the definition of the two parameters connected by $\rho_0$ through equation (\ref{mhalocrit}).

\begin{figure} 
\begin{center}
\includegraphics[width=60mm]{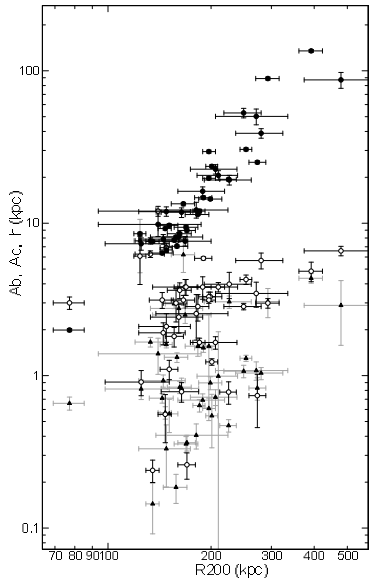} 
\end{center}
\caption{ Size-Size relations for dark halo critical radius for $(\rhalo,\ab)$ (triangles); $(\rhalo,\ad)$ (open circles); and  $(\rhalo,h)$ (black dots) (Sofue 2016). }  
\label{ssrelation}  
\end{figure}   

The figure also shows $\mbd$ plotted against $\mhalo$, which may be compared with the relation of stellar masses of galaxies against dark halo masses as obtained by cosmological simulation of star formation and hierarchical structure formation by Behroozi et al. (2013). The grey dashed line shows the result of simulation at $z=0.1$.

\begin{figure} 
\begin{center}
      \includegraphics[width=70mm]{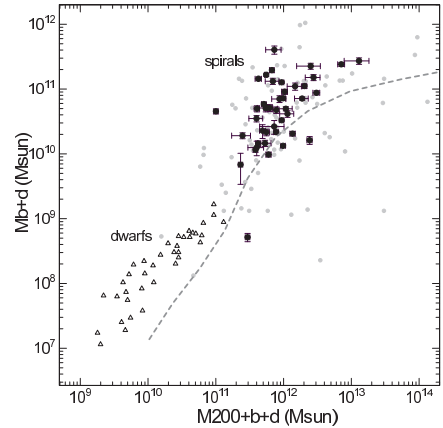}   
\end{center}
\caption{$\mbd$-$M_{\rm 200+b+d}$ relation compared with the stellar mass-total mass relation for dwarf galaxies (triangle: Miller et al. 2014) and simulation + photometry (grey dashed line: Behroozi et al. 2013). Black dots are the selected galaxies with reasonable fitting results, while small grey dots (including black dots) show non-weighted results from automatic decomposition of all rotation curves.} 
\label{mmdwsp} 
\end{figure}  

Figure \ref{mmdwsp} shows  the bulge+disk mass, $\mbd$, plotted against total mass, $M_{\rm 200+b+d}=\mhalo+\mbd$. In the figure we also plot photometric luminous mass and Virial mass obtained for dwarf galaxies by Millar et al. (2014). Also compared is a cosmological simulation by Behroozi et al. (2013). The simulation is in agreement in its shape with the plots for spiral and dwarf galaxies.  
The shape of simulated relation is consistent with the present dynamical observations, while the absolute values of $\mbd$ are greater than the simulated values by a factor of three. Solving the discrepancy may refine the cosmological models and will be a subject for the future.


The obtained correlations between the size and mass are the dynamical manifestation of the well established luminosity-size relation by optical and infrared photometry (de Jong et al. 1999; Graham and Worley 2008; Simard et al. 2011). It is interesting to notice that a similar size-mass relation is found for dark halos.  
  
The size-mass correlations can be fitted by straight lines on the log-log planes using the least-squares fitting. Measuring the mass and scale radii in $\Msun$ and kpc, respectively, we have size-mass relations for the  disk and dark halo as 
\be
\log  \md  = (9.89\pm      0.23)+(1.38\pm      0.41) ~\log \ad  ,
\label{logmdad}
\ee 
\be
\log  \mh   = (9.26\pm      0.52)+(1.45\pm      0.43) ~\log   h .
\label{logmhh}
\ee    
However, the relation for the bulge is too diverged to be fitted by a line.

The size-mass relation for the  disk agrees with the luminosity-size relation obtained by Simard et al. (2011). It is found that the relations for the disk and dark halo can be represented by a single (common) relation
\be
\log M_i=    (10.18\pm      0.24) +   (1.38\pm      0.21) \log a_i,
\label{fitmbdmhalo}
\ee
where $M_i=\mbd$ or $\mhalo$ in $\mten$ and $a_i=\ad$ or $h$ in kpc.
This simple equation leads to a relation between the bulge+disk mass to halo mass ratio, which approximately represents the baryonic fraction, expressed by the ratio of the scale radii of disk to halo as 
\be
\mbd/\mhalo\simeq (\ad/h)^{1.38}.
\ee
For the mean values of $<\ad>=3.3$ kpc and $<h>=21.6$ kpc, we obtain $\mbd/\mhalo \sim 0.07$.  
    
  		\chapter{CONCLUDING REMARKS}
                
                \section{Summary}
 
We reviewed the methods to observe and determine rotation curves of the Milky Way and spiral galaxies. 
Rotation characteristics of spiral galaxies and mass distributions were shown to be similar among the various types of galaxies. The dynamical structures are, thus, universal from Sa to Sc galaxies. The Milky Way exhibits representative universal characteristics from the central black hole to the dark halo. 
 
Major methods to derive the dynamical mass distribution in disk galaxies were described, and some results were presented. 
The direct method does not employ any galactic models or functional forms, but straightly compute the mass distribution. On the other hand, the decomposition method yields fundamental parameters representing the bulge, disk and dark halo individually. The decomposition method is thus convenient to discuss the fundamental structures composed of bulge, disk and dark halo. The fitted dynamical parameters were also useful for statistical correlation analyses such as the size-mass and mass-mass relations, which provides with fundamental information for formation and evolution scenarios of galaxies in the universe.

\section{Achievements in the Decades and the Future}

In our previous review (Sofue and Rubin 2001) various advances have been expected for the future rotation curve studies thanks principally to instrumental developments. We summarize and evaluate the progress in the decades.

\subsection{Extinction-free rotation kinematics in the central regions}

This has been achieved extensively using CO line spectroscopy and imaging, and will be achieved in an ultimate way using the ALMA (Atacama Large mm and submm Array) with higher spatial (0.01 arcsec) resolution. 
Spectroscoy with eight- and ten-meter class optical/infrared telescopes using infrared spectral lines such as Br $\gamma$ was anticipated, while the progress seems not fast, and no outstanding advances have been achieved. 

\subsection{VLBI astrometry} 

The VERA (VLBI Experiments for Radio Astrometry) observations have successfully achieved high-resolution astrometric measurements of paralactic distances, positions, proper motions, and radial velocities for a number of maser sources in the Milky Way. The observations resulted in a series of special volume in PASJ (Honma et al. 2015, and the literature therein). Many new aspects have been learnt in details of the three-dimensional kinematics of the Galaxy, which include the new values of $(R_0,V_0)$, non-circular motions related to the density waves and bars. The SKA (Square-Kilometer Array) will be a future, probably more powerful tool for trigonometric determination of the Galactic rotation curve and dynamics even by simply extending the method employed with VERA. 

\subsection{High-redshift rotation curves}
Thanks to the increase of aperture and sensitivity of the telescopes, particularly using the Hubble Space Telescope and ground-based 8-10 meter telescopes, galaxies at cosmological distances are becoming routine targets for rotation curve observations. {\changed Tens of galaxies at high redshifts have their rotation velocities measured.} The data are still not sensitive enough to be compared with those for the nearby galaxies. The global rotation curve shapes seem to be similar already at $z\sim 3-5$. Combined with cosmological simulations of structural formation and evolution, this would become one of the most popular fields in rotation curve study. 

However, the inner rotation curves representing the bulge and central massive objects are still not resolved at high redshifts. As mentioned already, an angular resolution of $0''.01$ corresponds only to $\sim 200$ pc at $z>\sim 2$. Higher resolution instruments are desired for more detailed study of the bulge and black hole formation, which are one of the major events in the primeval galaxies.

\subsection{Method of analysis}
Rotation curves presented in this review have been basically derived by the popular methods like PV tracing and tilted ring methods. The PV iteration method was also applied, while only in a few cases.

Development of sophisticated iteration methods of 2D and 3D velocity fields to produce more accurate rotation curves has been anticipated. However, not much advance was obtained in the decades. These will lead to more tightly constrained mass deconvolution, and precise distributions of the stellar and dark matter masses.

\subsection{Dark halos}  
Dark halos are one of the major topics in the field not only of galaxy dynamics and kinematics but also of the structure formation and evolution in the expanding universe. 
As to the Milky Way and M31, it has been shown that their rotation curves, and hence the dark matter halos, seem to be merged at a half distance to each other. Both galaxies showed declining rotation at radii larger than $R\sim 50$ kpc. 
{Their outermost rotation curves are not flat anymore and is not well fitted by the current pseudo-isothermal models, but they seem to be better approximated by the NFW model. However, it may be mentioned that the isothermal model is  simple and is enough to represent dark halos up to $\sim 30-50$ kpc, beyond which it is still difficult to obtain rotation curves in usual galaxies. 

Thus, the measurements of the dynamical mass of dark halos in most galaxies are limited to their outermost radii of measured rotation curves only to $\sim 20-30$ kpc, where it is difficult to discriminate the halo models. {\changed Measurements to rato larger radii, up to $\sim 100$ kpc,} for a more number of galaxies are crucial for conclusive comparison with the cosmological scenarios of structural formation. }

Dark matter density in the solar neighborhood has been determined in the range around $\sim 0.2-0.4{\rm GeV~cm}^{-3}$. 
The physical property of the dark matter in view of elementary particle physics would be clarified when direct detection in the laboratory is achieved. Also indirect detection toward the Sun and the Galactic Center are proposed and partly performed. Although, at the moment, there appears no report of firm detection, this would be the fundamental field in the experimental physics in the near future. 

\subsection{Massive black holes}
Coevolution of the spheroidal component and central supermassive black hole is a standard scenario in the structural evolution of galaxies based on the observational facts that black hole mass is positively correlated with the spheroidal mass in measured galaxies. This topic was, however, not reviewed in this paper, mainly because of insufficient sample galaxies with both the black hole mass and detailed rotation curves around the nuclei. Such study would be a promising subject for high-resolution submillimeter spectroscopy using ALMA and infrared spectroscopy with large-aperture telescopes including TMT.

\subsection{Activities }
Many of the galaxies with well defined rotation curves as discussed in this review are known as galaxies possessing various activities such as starburst, outflow, jets, and/or active galactic nuclei (AGN). However, there has been no clear correlation analysis among these activities and the galactic mass distribution. It could be a subject for a more detailed comparison study with careful inspection of the individual curves and mass structures, categorizing the curves into those with and without activities. It may be related to central bars by which the gas is accumulated to the nuclei, and hence the correlation of rotation curves of bar and non-barred galaxies.

\subsection{Star formation}
Correlation analysis of general star forming activity in the disk with the rotation curves and mass structure would be a fundamental subject related to ISM physics in galaxies. A relation is known that low surface brightness galaxies are generally slowly rotating and less massive. On the other hand, high velocity rotators, usually of late types as Sa and S0, show that their major star formation activity was over in the far past. In Sb and Sc galaxies, the star formation rate is not necessarily controlled by the mass structure, but rather related to spiral arms and/or bars, or to environmental effect such as interaction with other galaxies.

The relation between mass structure and star forming activity is, therefore, not well studied from observations of galaxies at the present epoch (redshift $z\simeq 0$). This is readily shown by the observed constant M/L ratio among galaxies, where M/L ratios do not represent star forming activity. Direct relation between the mass structure and star formation would be studied by rotation curve analysis of high-redshift galaxies during their major star formation.

\subsection{Bars and spiral arms} 
{\changed In the current rotation curve studies the bar/non-bar discrimination has not necessarily been the major subject for the reason raised in chapter 2. In this article we reviewed on rotation curves based on the assumption that the galactic rotation is circular and axisymmetric. Therefore, we did not touch upon spiral arms and bars in detail. 

The separation of arm and bar related parameters is still difficult even though the data get wider, deeper, and more accurate. The number of parameters, hence the number of freedom, is still large to reach a unique result. The difficulty has not been solved even by numerical simulations.

Some new technique to directly analyze the mass distribution and motion in arms and bars are desired. More detailed inspection, probably in a more sophisticated way, of spectroscopic 2D and 3D data cubes of a larger number of galaxies, particularly data with such facilities like 2D Fabry-Perot instruments, velocity cubes in HI and CO using more advanced interferometers like SKA and ALMA, would lead to deeper insight into mass structures of galaxies including bars and spiral arms.}
\\

\noindent{\bf References}
\\    
 
  
\def\r{\hangindent=1pc  \noindent}   
              
\r Aaronson M, Mould J. 1986.   Ap. J. 303:1  

\r Amram P, Marcelin M, Balkowski C, Cayatte V, Sullivan III WT, Le Coarer E. 1994.   Astron. Astrophys. Supp.  103:5     

\r Arimoto, N., Sofue, Y., and Tsujimoto, T. 1996, PASJ, 48, 275

\r   Battinelli, P., Demers, S., Rossi, C.,  and  Gigoyan, K.~S.\ 2013, Astrophysics, 56, 68 

\r {\changed Athanassoula, E.\ 1992, \mnras, 259, 328 }

\r Begeman KG, Broeils AH, Sanders RH. 1991. MNRAS 249:523 

\r Begeman KG. 1989 AA 223:47   

\r Behroozi, P.~S., 
Wechsler, R.~H., \& Conroy, C.\ 2013, \apj, 770, 57 
  
\r Bershady, M.~A., Verheijen, M.~A.~W., Westfall, K.~B., et al.\ 2010a, \apj, 716, 234 

\r  Bershady, M.~A., Verheijen, M.~A.~W., Swaters, R.~A., et al.\ 2010b, \apj, 716, 198  
  
\r Bertola F, Cappellari M, Funes JG, Corsini EM, Pizzella A, Vega Bertran JC. 1998.   Ap. J. Lett.  509:93   
 
\r Bhattacharjee, P., Chaudhury, S., Kundu, S.,  and  Majumdar, S.\ 2013, \prd, 87, 083525 

\r Bhattacharjee, P., Chaudhury, S.,  and  Kundu, S.\ 2014, \apj, 785, 63 
 
\r  Binney, J.,  and  Dehnen, W.\ 1997, \mnras, 287, L5 

\r Binney J, Gerhard OE, Stark AA, Bally J, Uchida KI 1991 MNRAS 252:210  

\r Binney J, Tremaine S. 1987, in   Galactic Dynamics . Princeton Univ. Press

\r Blais-Ouellette, S., Amram, P., Carignan, C. 2001 AJ 121, 1952 

\r Blais-Ouellette, S., Amram, P., Carignan, C.,  and  Swaters, R.\ 2004, \aap, 420, 147 
 
\r Blitz, L. 1979.   Ap. J.  227:152 

\r Blitz, L., Fich, M., Stark, AA 1982, ApJs, 49, 183    

\r Bobylev, V.~V.\ 2013, Astronomy Letters, 39, 95  

\r Bobylev, V.~V., \& Bajkova, A.~T.\ 2015, Astronomy Letters, 41, 473 
 
\r Bosma A. 1981a. AJ 86:1825  

\r Bosma A. 1981b. AJ 86:1791    

\r Bovy, J., \& Tremaine, S.\ 2012, \apj, 756, 89 

\r Broeils AH. 1992.   Astron. Astrophys . 256:19
  
\r Burkert, A., 1995, ApJ, 447, L25  

\r Burton, W. B., Gordon, M. A. 1978 AA 63, 7. 

\r Burton, W. B., Liszt, H. S. 1993 AA 274, 765. 

\r Buta R, Purcell GB, Cobb ML, Crocker DA, Rautiainen P, Salo H. 1999.   Astron. J.  117:778 

\r   Carignan, C.\ 1985, \apj, 299, 59  

\r Carignan C, Beaulieu S. 1989.   Ap. J.  347:760

\r Carignan C, Freeman KC. 1985.   Ap. J.  294:494 

\r Carignan C, Puche D. 1990a. AJ 100:394

\r Carignan C, Puche D. 1990b. AJ 100:641   

\r Carignan, C., Chemin, L., Huchtmeier, W.~K.,  and  Lockman, F.~J.\ 2006, \apjl, 641, L109

\r Catinella, B., Giovanelli, R.,  and  Haynes, M.~P.\ 2006, \apj, 640, 751 

\r  Chemin, L., Carignan, C.,  and  Foster, T.\ 2009, \apj, 705, 1395 
, \aap, 511, A89 

\r Ciotti L., 1991, AA, 249, 99.

\r Clemens, D. P. 1985. ApJ 295:422    

\r Corradi RLM, Boulesteix J, Bosma A, Amram P, Capaccioli M. 1991.   Astron. Astrophys.  244:27 

\r Courteau, S. 1997. AJ 114:2402

\r Dicaire et al. 2008, MNRAS, 385, 553

\r Daigle, O., Carignan, C., Amram, P., et al.\ 2006, \mnras, 367, 469 

\r de Blok WJG, McGaugh SS, van der Hulst JM. 1996.   MNRAS  283:18 

\r de Blok, W. J. G. 2005 ApJ 634 227 

\r de Blok, W.~J.~G., McGaugh, S.~S.,  and  Rubin, V.~C.\ 2001, \aj, 122, 2396 

\r de Blok, W.~J.~G.,  and  Bosma, A.\ 2002, \aap, 385, 816 

\r de Blok, W.~J.~G., Walter, F., Brinks, E., et al.\ 2008, \aj, 136, 2648 

\r de Vaucoureurs, G. 1958, ApJ, 128, 465  

\r Dehnen, W.,  and  Binney, J.\ 1998, \mnras, 294, 429  

\r Demers, S., Battinelli, P. 2007 AA, 473, 143.  

\r Di Teodoro, E.~M., \& Fraternali, F.\ 2015, \mnras, 451, 3021 
 
\r Dicaire, I., Carignan, C., Amram, P., et al.\ 2008, \mnras, 385, 553  

\r  Epinat, B., Tasca, L., Amram, P., et al.\ 2012, \aap, 539, AA92 

\r  Erb, D.~K., Shapley, A.~E., 
Steidel, C.~C., et al.\ 2003, \apj, 591, 101 
  
\r Erroz-Ferrer S., et al., 2012, MNRAS, 427, 2938 

\r Feast, M.,  and  Whitelock, P.\ 1997, \mnras, 291, 683  

\r Ferrarese L, Ford HC. 1999.   Ap. J.  515:583 
 
\r Fich, M., Blitz, L., Stark, A. A. 1989 ApJ 342, 272  

\r Fich, M,, Tremaine, S. 1991 ARAA, 29, 409 
 
\r Forbes, DA. 1992 AA Suppl. 92:583   

\r Freeman, K. C. 1970, ApJ, 160, 811   
  
\r Garrido, O., Marcelin, M., Amram, P.,  and  Boulesteix, J.\ 2002, \aap, 387, 821  

\r Garrido, O., Marcelin, M.,  and  Amram, P.\ 2004, \mnras, 349, 225 

\r Garrido, O., Marcelin, M., Amram, P., et al.\ 2005, \mnras, 362, 127 

\r  Gentile G., Salucci P., Klein U., Granato G.~L., 2007, MNRAS, 375, 199 

\r  Gentile G., et al., 2015, A\&A, 576, A57 
 
\r Genzel, R., Hollenbach, D.,  and  Townes, C.~H.\ 1994, Reports on Progress in Physics, 57, 417 

\r Genzel R, Eckart A, Ott T, Eisenhauer F. 1997. MNRAS 291:219   

\r Genzel, R., Pichon, C., Eckart, A., Gerhard, O.~E.,  and  Ott, T.\ 2000, \mnras, 317, 348 

\r Genzel, R., Thatte, N., Krabbe, A., Kroker, H., Tacconi-Garman, L 1996 ApJ 472, 153 

\r  Genzel, R., Burkert, A., Bouch{\'e}, N., et al.\ 2008, \apj, 687, 59 

\r Genzel, R., Eisenhauer, F.,  and  Gillessen, S.\ 2010, Reviews of Modern Physics, 82, 3121 

\r  Genzel, R., Newman, S., Jones, T., et al.\ 2011, \apj, 733, 101  

\r Ghez A, Morris M, Klein BL, Becklin EE. 1998. ApJ 509:678.  

\r Ghez, A. M., Salim, S., Hornstein, S. D., Tanner, A., Lu, J. R., Morris, M., Becklin, E. E., Duchene, G. 2005 ApJ 620, 744.  

\r Ghez, A.~M., Salim, S., Weinberg, N.~N., et al.\ 2008, \apj, 689, 1044 

\r Gillessen, S., Eisenhauer, F., Trippe, S., et al.\ 2009, \apj, 692, 1075   

\r  Graham, A.~W., \& Worley, C.~C.\ 2008, \mnras, 388, 1708 

\r Greenhill LJ, Gwinn CR, Antonucci R, Barvainis R. 1996.   Ap. J. Lett.  472:L21 

\r H\'eraudeau Ph, Simien F. 1997.   Astron. Astrophys . 326:897 

\r Haschick AD, Baan WA, Schneps MH, Reid MJ, Moran JM, Guesten R. 1990.   Ap. J. , 356:149. 
\r  Hernandez, O., Carignan, C., Amram, P., Chemin, L., \& Daigle, O.\ 2005, \mnras, 360, 1201 

\r Herrnstein JR, Moran JM, Greenhill LJ, Diamond PJ, Inoue M, Nakai N, Miyoshi M, Henkel C, Riess A. 1999. Nature 400:539   
 
\r Hlavacek-Larrondo J., Marcelin M., Epinat B., Carignan C., de Denus-Baillargeon M.-M., Daigle O., Hernandez O., 2011a, 
MNRAS, 416, 509 

\r  Hlavacek-Larrondo J., Carignan C., Daigle O., de Denus-Baillargeon M.-M., Marcelin M., Epinat B., Hernandez O., 2011b, 
MNRAS, 411, 71 

\r Honma M, Sofue Y. 1996.   Publ. Astron. Soc. Japan Lett.  48:103 

\r Honma M, Sofue Y. 1997a.   Publ. Astron. Soc. Japan.  49:453 

\r Honma M, Sofue Y. 1997b.   Publ. Astron. Soc. Japan.  49:539 

\r  Honma, M., Bushimata, T., 
Choi, Y.~K., et al.\ 2007, \pasj, 59, 889   

\r  Honma, M., Nagayama, T., Ando, K., et al.\ 2012, \pasj, 64, 136  

\r Honma, M., Nagayama, T.,  Sakai, N.\ 2015, \pasj, 67, 70  

\r Hunter JH, Gottesman ST. 1996. in   Barred Galaxies . eds. R Buta, DA Crocker, Elmegreen BG.   PASP. Conf. Series . 91:398  
 
\r Irwin Judith A, Sofue Y. 1992.   Ap. J. Lett.  396:L75.   

\r Jenkins, A.,  and  Binney, J.\ 1994, \mnras, 270, 703  imulation
  
\r Jobin M, Carignan C. 1990. AJ 100:648  

\r Kamphuis J, Briggs F. 1992.   Astron. Astrophys.  253:335 

\r Kelson DD, Illingworth GD, van Dokkum PG, Franx M. 2000   Ap. J.   

\r Kent, S. M. 1986, AJ, 91, 1301 

\r Kent, S. M. 1992. ApJ 387:181

\r Kent, S. M., Dame, T.~M.,  and  Fazio, G.\ 1991, ApJ, 378, 131  

\r Kerr, F.~J.,  and  Lynden-Bell, D.\ 1986, \mnras, 221, 1023  

\r Kim S, Stavely-Smith L, Dopita MA, Freeman KC. Sault RJ, Kesteven MJ, McConnell D. 1998.   Ap. J.  503:674.  

\r Koda J, Sofue Y, Wada K. 2000a.   Ap. J.  532:214

\r Koda J, Sofue Y, Wada K. 2000b.   Ap. J. L.  531:L17
 
\r Kong, D. L.; Zhu, Z. 2008, Acta Astronomica Sinica, 49, 224  

\r Kormendy J, Richstone D. 1995 ARAA 33:581 

\r Kormendy J, Westpfahl DJ, 1989. ApJ 338:752   

\r Kormendy J. 2001. in   Galaxy Disks and Disk Galaxies . eds. J Funes, E Corsini.   PASP. Conf. Series , in press 

\r Krabbe A, Colina L, Thatte N, Kroker H. 1997.   Ap. J.  476:98 

\r Kuno N, Nishiyama K, Nakai N, Sorai K, Vila-Vilaro B. 2000 PASJ 52, 775  

\r Kwee, K. K., Muller, C. A., and Westerhout, G. BAN 1954, 12, 211.  

\r Lake G, Schommer RA, van Gorkom JH. 1990.   Astron. J.  99:547  

\r Law, D.~R., Steidel, C.~C., 
Erb, D.~K., et al.\ 2009, \apj, 697, 2057  

\r Lindqvist, M., Habing, H.~J.,  and  Winnberg, A.\ 1992, \aap, 259, 118  

\r  L{\'o}pez-Corredoira, M.\ 2014, \aap, 563, A128   

\r Makarov, D.~I., Karachentsev, I.~D., Tyurina, N.~V., 
 and  Kaisin, S.~S.\ 1997, Astronomy Letters, 23, 445 

\r Makarov, D.~I., Burenkov, A.~N.,  and  Tyurina, N.~V.\ 2001, Astronomy Letters, 27, 213

\r  M{\'a}rquez I., Masegosa J., Moles M., Varela J., Bettoni D., Galletta G., 2002, A\&A, 393, 389  

\r Martinsson, T.~P.~K., Verheijen, M.~A.~W., Westfall, K.~B., et al.\ 2013, \aap, 557, AA131 
 
\r Mathewson DS, Ford VL, Buchhorn M. 1992 ApJ Suppl 81:413 

\r Mathewson DS, Ford VL. 1996 ApJ Supp 107:97   

\r McGaugh, S.~S., Rubin, V.~C.,  and  de Blok, W.~J.~G.\ 2001, \aj, 122, 2381 
\r McGaugh SS,  and  de Blok WJG. 1998.   Ap. J.  499:66 

\r Merrifield M. R. 1992. AJ, 103, 1552.  

\r  Mignard, F.\ 2000, \aap, 354, 522  

\r  Masters, K.~L., Crook, A., Hong, T., et al.\ 2014, \mnras, 443, 1044 

\r  Masters, K.~L., Springob, C.~M., \& Huchra, J.~P.\ 2008, \aj, 135, 1738   

\r Miller S.~H., Ellis R.~S., Newman A.~B., Benson A., 2014, ApJ, 782, 115  

\r Miyoshi M, Moran J, Herrnstein J, Greenhill L, Nakai N, Diamond P, Inoue M. 1995 Nature 373:127.   

\r Nakai N, Inoue M, Miyoshi M 1993. Nature 361:45  

\r Nakai, N. 1992 PASJ, 44, L27  

\r Nakanishi, H., Sofue, Y. 2003 PASJ 55, 191.

\r Nakanishi, H., Sofue, Y.  2006 PASJ 58, 847. 

\r  Nakanishi, H., \& Sofue, Y.\ 2016, \pasj, 68, 5

\r Nakanishi, H., Sakai, N., Kurayama, T., et al.\ 2015, \pasj, 67, 68 

\r Navarro, J. F., Frenk, C. S., White, S. D. M., 1996, ApJ, 462, 563    
M., Tully, R.~B., et al.\ 2014, \apj, 792, 129   

\r Navarro, J.~F., Frenk, 
C.~S., \& White, S.~D.~M.\ 1997, \apj, 490, 493 

\r Noordermeer, E., van der Hulst, J.~M., Sancisi, R., Swaters, R.~A., \& van Albada, T.~S.\ 2005, \aap, 442, 137  

\r Noordermeer, E.\ 2008, MNRAS, 385, 1359 

\r   Noordermeer, E., van der Hulst, J.~M., Sancisi, R., Swaters, R.~S.,  and  van Albada, T.~S.\ 2007, \mnras, 376, 1513  

\r   Oikawa, S., \& Sofue, Y.\ 2014, \pasj, 66, 77 

\r Oka T, Hasegawa T, Sato F, Tsuboi M, Miyazaki A, 1998. ApJ S 118:455   

\r Olling R.~P., 1996, AJ, 112, 457 

\r Olling, R.~P., \& Merrifield, M.~R.\ 1998, \mnras, 297, 943 

\r  Olling, R.~P., \& Dehnen, W.\ 2003, \apj, 599, 275 

\r Oort J. H. 1965 in   Stars and Stellar Systems, Vol. 5, Galactic Structure , ed. A. Blaauw and M. Schmidt (Univ. Chicago Press), p. 455. 

\r Oort, J. H. and Kerr, F. J. 1958 MNRAS, 114, 3790.

\r Oort, J. H., Kerr, F. J., Westerhout, G  1958 MNRAS 118, 379. 

\r  Pato, M., Iocco, F., \& Bertone, G.\ 2015a, JCAP, 12, 001 

\r  Pato, M., \& Iocco, F.\ 2015b, \apjl, 803, L3 
 
\r Persic M, Salucci P. 1990   MNRAS . 247:349 

\r  Persic, M.,  and  Salucci, P.\ 1991, \apj, 368, 60  

\r Persic M, Salucci P, Stel F. 1996. MNRAS 281:27

\r  Piffl, T., Binney, J., McMillan, P.~J., et al.\ 2014, \mnras, 445, 3133 

\r Puche D, Carignan C, Bosma A. 1990.   Astron. J.  100:1468

\r Puche D, Carignan C, Wainscoat RJ. 1991a.   Astron. J.  101:447

\r Puche D, Carignan C, van Gorkom JH. 1991b.   Astron. J.  101:456

\r  Randriamampandry, T.~H., Combes, F., Carignan, C., \& Deg, N.\ 2015, \mnras, 454, 3743

\r Regan MW, Vogel SN. 1994.   Ap. J.  43:536

\r Reid, M. J. 1993, ARAA, 31, 345   

\r Reid, M.~J., Menten, K.~M., Zheng, X.~W., et al.\ 2009, \apj, 700, 137-148  

\r Reyes R., Mandelbaum R., Gunn J.~E., Nakajima R., Seljak U., Hirata C.~M., 2012, MNRAS, 425, 2610 

\r Richstone D, Bender R, Bower G, Dressler A, Faber S, et al. 1998.   Nature  395A:14 . 

\r Rieke, G.~H.,  and  Rieke, M.~J.\ 1988, \apjl, 330, L33 
 
\r   Roberts, M.~S.\ 1966, \apj, 144, 639  

\r   Roeser, S., Demleitner, M., \& Schilbach, E.\ 2010, \aj, 139, 2440  

\r Robertson, B.~E., \& Bullock, J.~S.\ 2008, \apjl, 685, L27 
 
\r Roscoe DF. 1999 AA 343:788  

\r  Rots, A.~H., Bosma, A., van der Hulst, J.~M., Athanassoula, E.,  and  Crane, P.~C.\ 1990, \aj, 100, 387   
 
\r Rubin VC, Burstein D, Ford Jr WK, Thonnard N. 1985. ApJ 289:81  

\r Rubin VC, Ford Jr WK, Thonnard N. 1982. ApJ 261:439    

\r Rubin VC, Hunter DA, Ford Jr WK. 1991 ApJ Supp 76:153    

\r Rubin VC, Kenny JDP, Young, JS. 1997. AJ 113:1250 

\r Rubin VC, Thonnard N, Ford Jr WK. 1982. AJ 87:477   

\r Rubin VC, Waterman AH, Kenney JD, 1999. AJ 118:236   

\r  Ryder S.~D., Zasov A.~V., McIntyre V.~J., Walsh W., Sil'chenko O.~K., 1998, MNRAS, 293, 411 

\r  Sakai, N., Honma, M., Nakanishi, H., et al.\ 2012, \pasj, 64, 108 

\r Sakai, N., Nakanishi, H., Matsuo, M., et al.\ 2015, \pasj, 67, 69  

\r Salucci P, Ashman KM, Persic M. 1991. ApJ 379:89  

\r Salucci P, Frenk CS. 1989.   MNRAS . 237:247

\r   Salucci, P., Lapi, A., Tonini, C., et al.\ 2007, \mnras, 378, 41 

\r Salucci, P., Nesti, F., Gentile, G., Martins, C. F. 2010 AA 523,83.
  
\r Sawada-Satoh S, Inoue M, Shibata KM, Kameno S, Migenes V, Nakai N, Diamond PJ. 2000.   Publ. Astron. Soc. Jpn.  52:421 
 
\r  Shetty, R., Vogel, S.~N., Ostriker, E.~C.,  and  Teuben, P.~J.\ 2007, \apj, 665, 1138  

\r Schombert JM, Bothun GD, Schneider SE, McGaugh SS. 1992 AJ  103:1107

\r Schombert JM. Bothun GD. 1988.   Astron. J.  95:1389 

\r Shapiro, K.~L., Genzel, 
R., F{\"o}rster Schreiber, N.~M., et al.\ 2008, \apj, 682, 231 

\r Shlosman I, Begelman MC, Frank, J 1990.   Nature . 345:679 

\r Sofue, Y. 1995a, \pasj, 47, 551  

\r Sofue Y. 1995b PASJ 47:527   

\r Sofue, Y. 1996, ApJ 458, 120 

\r Sofue Y. 1997. PASJ 49:17 

\r Sofue Y. 1998. PASJ 50:227 

\r Sofue Y. 1999. PASJ 51:445 

\r Sofue, Y. 2009, PASJ, 61, 153 

\r Sofue, Y. 2011, \pasj, 63, 813 

\r Sofue, Y. 2012, \pasj, 64, 75   

\r Sofue, Y. 2013a, in "Planets, Stars and Stellar Systems", Springer, Vol. 5, Chap. 19,  ed. G. Gilmore.

\r Sofue, Y. 2013b, \pasj, 65, 118  

\r  Sofue, Y. 2015a, \pasj, 67, 75

\r   Sofue, Y.\ 2016, \pasj, 68, 2  

\r Sofue, Y., Honma, M., Omodaka, T. 2009 PASJ, 61, 227. 

\r Sofue Y, Irwin Judith A. 1992.   Publ. Astron. Soc. Jpn.  44:353 

\r Sofue Y, Koda J, Kohno K, Okumura SK, Honma M, Kawamura A, Irwin J A. 2001 ApJL 547, L115  

\r  Sofue Y., Koda J., Nakanishi H., Onodera S., 2003, PASJ, 55, 59 

\r Sofue, Y. and Nakanishi, H. 2016 PASJ, in press.

\r Sofue, Y., Rubin, V. C. 2001 ARAA 39, 137    

\r Sofue Y, Tomita A, Honma M, Tutui Y, Takeda Y. 1998. PASJ  50:427

\r Sofue Y, Tomita A, Honma M, Tutui Y. 1999b. PASJ 51:737 

\r Sofue Y, Tutui Y, Honma M, Tomita A. 1997.   Astron. J.  114:2428

\r Sofue Y, Tutui Y, Honma M, Tomita A, Takamiya T, Koda J, Takeda Y. 1999a. ApJ 523:136   
 
\r  Spano, M., Marcelin, M., Amram, P., et al.\ 2008, \mnras, 383, 297  

\r  Spekkens, K., \& Sellwood, J.~A.\ 2008, ApSpSci, Proceedings, 5, 167 

\r Spergel, D. N., et al. 2003, ApJs, 148, 175    

\r Swaters RA, Madore BF, Trewhella M. 2000.   Ap. J. Lett.  356:L49. 

\r Swaters, R.~A., \& Balcells, M.\ 2002, \aap, 390, 863  

\r Swaters, R.~A., Sancisi, R., van Albada, T.~S.,  and  van der Hulst, J.~M.\ 2009, \aap, 493, 871

\r Simard, L., Mendel, J.~T., Patton, D.~R., Ellison, S.~L., \& McConnachie, A.~W.\ 2011, \apjs, 196, 11 
  
\r Takase, B. 1957 PASJ 9, 16.

\r Takamiya T, Sofue Y. 2000. ApJ 534:670   

\r Takamiya, T.,  and  Sofue, Y.\ 2002, \apjl, 576, L15 

\r Tecza M, Thatte N, Maiolino R. 2000. IAU Symposium No. 205.   Galaxies and their Constituents at the Highest Angular Resolution  in press   

\r Trotter AS, Greenhill LJ, Moran JM, Reid MJ, Irwin JudithA, Lo K-Y. 1998.   Ap. J.  495:740 

\r Trujillo, I., et al. 2002, MNRAS, 333, 510   

\r Tully RB, Fisher JR. 1977.   Astron. Astrophys . 54:661 
  
\r van der Marel RP, Rix HW, Carter D, Franx M, White SDM, de Zeeuw T. 1994.   MNRAS . 268:521 

\r van der Wel, A., \& van der Marel, R.~P.\ 2008, \apj, 684, 260 

\r van Gorkom, J.~H., van der Hulst, J.~M., Haschick, A.~D., \& Tubbs, A.~D.\ 1990, \aj, 99, 1781 

\r Vaughan JM. 1989.   The Fabry-Perot interferometer. History, theory, practice and applications.  The Adam Hilger Series on Optics and Optoelectronics, Bristol: Hilger. 

\r Vogel, S.~N., Rand, R.~J., Gruendl, R.~A.,  and  Teuben, P.~J.\ 1993, \pasp, 105, 666 
 
\r Vogt NP, Forbes DA, Phillips AC, Gronwall C, Faber SM, Illingworth GD, Koo DC. 1996.   Ap. J. Lett.  465:15

\r Vogt NP, Herter T, Haynes MP, Courteau S. 1993.   Ap. J. Lett.  415:95

\r Vogt NP. Phillips AC, Faber SM, Gallego J, Gronwall C, Guzman R, Illingworth GD, Koo DC, Lowenthal J D. et al. 1997.   Ap. J. Lett.  479:121

\r Vogt, N.~P., Haynes, M.~P., Herter, T., \& Giovanelli, R.\ 2004a, \aj, 127, 3273  

\r Vogt, N.~P., Haynes, M.~P., Giovanelli, R., \& Herter, T.\ 2004b, \aj, 127, 3325  

\r Watson WD, Wallim BK. 1994.   Ap. J.  432:35 

\r Weber, M. and de Boer, W. 2010 AA 509, 25.  

\r Weiner BJ, Williams TB. 1996.   Astron. J.  111:1156
 
\r Westfall, K.~B., Andersen, D.~R., Bershady, M.~A., et al.\ 2014, \apj, 785, 43 

\r Whitmore B.~C., McElroy D.~B., Schweizer F., 1987, ApJ, 314, 439 
 

\end{document}